\documentclass[prd,twocolumn,superscriptaddress,amsfonts,amssymb,amsmath,showpacs]{revtex4-2}
\usepackage{amssymb}
\usepackage{bm}
\usepackage{amsfonts}
\usepackage{latexsym}
\usepackage[latin1]{inputenc}
\usepackage{graphicx}
\usepackage{amsmath}
\usepackage{palatino}
\usepackage{mathpazo}
\usepackage{textcomp}
\usepackage{float}
\usepackage{booktabs}
\usepackage{dcolumn}
\usepackage{ragged2e}
\usepackage{hyperref}
\hypersetup{colorlinks,citecolor=blue}
\hypersetup{colorlinks=true,
	linkcolor=purple,
	filecolor=magenta,
	citecolor=blue}
 
\usepackage[many]{tcolorbox}   
\newtcolorbox{boxB}{
    fontupper = \bf, 
    boxrule = 1.5pt,
     width = 18.8cm,
    colframe = black,
    colback = white,
    rounded corners,
    arc = 5pt   
    }

\usepackage{amsmath}
\usepackage{xcolor}
\usepackage{orcidlink}
\usepackage{epsfig}
\usepackage[caption=false]{subfig}
\usepackage{commath}
\usepackage{tabularx}
\usepackage{multirow}
\usepackage{mathtools}
\newcommand{\be}{\begin{equation}}
	\newcommand{\ee}{\end{equation}}
\newcommand{\bea}{\begin{eqnarray}}
	\newcommand{\eea}{\end{eqnarray}}
\newcommand{\beas}{\begin{eqnarray*}}
	\newcommand{\eeas}{\end{eqnarray*}}

\allowdisplaybreaks[1]
\renewcommand{\arraystretch}{1.1}
\begin{document}
	\color{black}       
	%
	\title{Modelling the Accelerating Universe with $f(Q)$ Gravity: Observational Consistency}
	\author{S.A. Narawade\orcidlink{0000-0002-8739-7412}}
	\email{shubhamn2616@gmail.com}
	\affiliation{Department of Mathematics, Birla Institute of Technology and
		Science-Pilani,\\ Hyderabad Campus, Hyderabad-500078, India.}
	\author{S. H. Shekh\orcidlink{0000-0003-4545-1975}}
	\email{da\_salim@rediff.com}
	\affiliation{Department of Mathematics, S.P.M. Science and Gilani Arts, Commerce College, Ghatanji, Yavatmal, Maharashtra 445301, India}
	\author{B. Mishra\orcidlink{0000-0001-5527-3565}}
	\email{bivu@hyderabad.bits-pilani.ac.in}
	\affiliation{Department of Mathematics, Birla Institute of Technology and
		Science-Pilani,\\ Hyderabad Campus, Hyderabad-500078, India.}
	\author{Wompherdeiki Khyllep\orcidlink{0000-0003-3930-4231}}
	\email{sjwomkhyllep@gmail.com}
	\affiliation{Department of Mathematics, St.\ Anthony's College, Shillong, Meghalaya 793001, India}
	\author{Jibitesh Dutta\orcidlink{0000-0002-6097-454X}}
	\email{jibitesh@nehu.ac.in}
	\affiliation{Mathematics Division, Department of Basic Sciences and Social Sciences, North-Eastern Hill University, Shillong, Meghalaya 793022, India.}
	\affiliation{Inter-University Centre for Astronomy and Astrophysics, Pune 411007, India}
	
\begin{abstract}
\textbf{Abstract:}
In this paper, we present a cosmological model within the framework of symmetric teleparallel gravity, focusing on $f(Q)$ gravity, where $Q$ represents the non-metricity scalar. Utilizing cosmological datasets, we derive an accelerating cosmological model by constraining its free parameters. To achieve this, we determine the parametric form of the Hubble parameter using a well-motivated $f(Q)$ function. Remarkably, all obtained values fall within the range suggested by cosmological observations. By employing the best-fit parameters, we calculate the present geometrical parameters and demonstrate the accelerating behaviour of the Universe. Furthermore, we thoroughly examine the evolutionary behaviours of the Universe, noting that our model converges to the $\Lambda$CDM model at late times. Finally, we investigate the energy conditions and find a violation of the strong energy condition, which could provide a valuable understanding of the nature of dark energy.

\end{abstract}

\maketitle
\section{Introduction}\label{Sec:I}
The theory of General Relativity (GR) stands as the most successful framework upon which numerous cosmological models have been proposed in scholarly works. It is crucial for each of these models to effectively account for the observed accelerated expansion of the Universe \cite{Riess1998, Perlmutter1999}. Within the context of GR, this accelerated expansion has been attributed to the presence of a peculiar form of energy known as Dark Energy (DE). The validity of this explanation has been strongly supported by highly precise observational evidence, including results from the Wilkinson Microwave Anisotropy Probe experiment (WMAP) \cite{Komatsu2003, Bennett2013, Komatsu2011, Hinshaw2013}, Baryonic Acoustic Oscillations (BAO) \cite{Eisenstein2005}, Large Scale Structure (LSS) \cite{Daniel2008}, Cosmic Microwave Background Radiation (CMBR) \cite{Fowler2010, Das2011, Keisler2011, Reichardt2012}, Baryon Oscillation Spectroscopic Survey (BOSS) \cite{Alam2016}, and the Planck Collaboration \cite{Ade2016, Aghanim2020, Aghanim2020a}.

Despite its success, DE models within the framework of GR have encountered numerous challenges. These challenges include the cosmological constant problem, which arises from the disparity between observed and predicted values of DE. Additionally, modified gravity and scalar field models face difficulties precisely in fine-tuning their parameters. Another significant issue is the coincidence problem, which raises doubts about the dominance of DE. Furthermore, the fundamental nature of DE remains unknown, leaving open questions regarding its origin, whether it arises from vacuum energy, modified gravity, or a dynamic scalar field \cite{Ruiz-Lapuente2010, Amendola2010}.

Due to the limitations of General Relativity (GR) in addressing late-time cosmic acceleration, modifications to the theory have become necessary. In the existing literature, two geometrically equivalent theories to GR are frequently discussed: the teleparallel equivalent of GR, based on torsion \cite{Aldrovandi2013, Maluf2001, Ferraro2008, Bengochea2009, Linder2010}, and the symmetric teleparallel equivalent of GR, based on non-metricity \cite{Nester1999}. Although these theories are equivalent to GR, their respective extensions differ fundamentally \cite{Altschul2015}.


In this study, we focus on the extension of symmetric teleparallel gravity, also known as $f(Q)$ gravity \cite{Jimenez2018}. In one of its formulations, sometimes referred to as coincident GR, this theory simplifies calculations by using specific coincident gauges, reducing the covariant derivative to a partial derivative. However, $f(Q)$ gravity also encompasses formulations with non-coincident gauges, which attract increased attention in cosmological investigations, whether in a flat Universe \cite{Subramaniam2023, Shabani2023, Paliathanasis2023} or a curved one \cite{Dimakis2022, Heisenberg2023, Shabani2023a, Subramaniam2023a}.

Recently, the $f(Q)$ theory of gravity has been successful in the cosmological phenomenology both at the background and perturbation levels \cite{Lu2019, Lazkoz2019, Jimenez2020, Ayuso2021,Hu2022, Esposito2022, Khyllep2021, Khyllep2023,Sahlu2022}. Also, it has been successful in confrontation with various observational data sets such as, CMBR, SNIa, BAO, redshift space distortion, etc. \cite{Soudi2019, Barros2020, Anagnostopoulos2021, Atayde2021, Frusciante2021}. Anagnostopoulos et al. \cite{Anagnostopoulos2023} showed that $f(Q)$ gravity can safely pass the Big Bang Nucleosynthesis constraints. The behaviour of the dynamical parameters of $f(Q)$ gravity constrained by the scale factor and model parameters was studied in Refs. \cite{Narawade2022a, Koussour2022a}. Maurya et al. \cite{Maurya2022} explored gravitationally decoupled anisotropic solutions for strange stars using the Complete Geometric Deformation (CGD) technique. In another line of research, $f(Q)$ theories have been proposed where the non-metricity $Q$ is non-minimally coupled either to the matter Lagrangian \cite{Harko2018} or to a scalar field \cite{Bahamonde2022}, leading to intriguing phenomenology. Furthermore, Capozziello et al. \cite{Capozziello2022} performed a reconstruction of the function $f(Q)$ using a numerical inversion procedure based on current observational constraints of cosmographic parameters. One can find some relevant research in symmetric teleparallel gravity and related to recent observations in Refs. \cite{Bajardi2020, Bajardi2021, Bajardi2023, Vagnozzi2020, Adil2023, Escamilla2023, Bernal2016, Adi2021, Odintsov2021,  Briffa2022}.

The $f(Q)$ gravity theory is part of symmetric teleparallel gravity, a modified framework that incorporates non-metricity. When considering perturbations around the Friedmann-Lema\^itre-Robertson-Walker (FLRW) background, this gravity theory does not present significant issues related to strong coupling. Therefore, the objective is to investigate the influence of this gravity theory on cosmological observations. Specifically, one can utilize expansion rate data obtained from cosmological observations to constrain the parameters involved in the parametrization of the redshift function, denoted as $H(z)$ \cite{Jimenez2018, Jarv2018, Harko2018, Runkla2018, Xu2019, Lazkoz2019}.

While the $f(Q)$ theory could challenge the $\Lambda$ Cold Dark Matter ($\Lambda$CDM) model \cite{Anagnostopoulos2021}, it is possible that phenomenological analyses of gravity models based on non-metricity may deviate from cosmological observations. Such deviations offer opportunities to differentiate these models from the standard $\Lambda$CDM scenario, making it essential to constrain the model parameters. Moreover, addressing the issue of late-time cosmic acceleration within this theoretical framework requires constraining the parameters using available data from cosmological observations. Recent studies employing this approach have yielded noteworthy results, as documented in references \cite{D'Ambrosio2022, Albuquerque2022, Narawade2022b, Koussour:2023rly}.

Motivated by these considerations, our focus will be to determine the parametric form of the Hubble parameter based on the functional form of $f(Q)$. The field equations of $f(Q)$ gravity will be derived using a spatially flat, isotropic, and homogeneous FLRW metric.

The paper is organized as: Sec. \ref{Sec:II} gives the basic mathematical formalism of $f(Q)$ gravity and the derivation of gravitational field equations are shown in Sec. \ref{Sec:III}. The $H(z)$ parametrization and other cosmographic parameters are shown in Sec. \ref{Sec:IV} and the observational constraints of the free parameters are obtained in Sec. \ref{Sec:V}. In Sec. \ref{Sec:VI} the cosmological tests for the validation of the model are studied. The analysis of the $f(Q)$ gravity model discussed in Sec. \ref{Sec:VII} and the conclusions are presented in Sec. \ref{Sec:VIII}.

\section{$f(Q)$ Gravity}\label{Sec:II}
In differential geometry, with the general connection $\Sigma_{\mu \nu }^{\alpha }$, the parallel transport of vectors and the computation of covariant derivatives can be performed. The metric tensor $g_{\mu \nu }$ provides information about angles, volumes, distances, and other geometrical properties. This can be thought of as a generalization of the gravitational potential in classical theory. In general, the connection $\Sigma_{\mu \nu }^{\alpha }$ can be decomposed into the following contributions, which include the torsion $T$, non-metricity $Q$, and curvature term $R$ \cite{Ortin2004}:

\begin{equation}\label{eq:1}
\Sigma_{\ \mu \nu }^{\alpha }=\Gamma_{\ \mu \nu }^{\alpha }+K_{\ \mu \nu}^{\alpha }+L_{\ \mu \nu }^{\alpha },
\end{equation}
where the famous \textit{Levi-Civita connection} $\Gamma _{\ \mu \nu }^{\alpha }$, which corresponds to the metric tensor $g_{\mu \nu }$, is given by: 
\begin{equation}\label{eq:2}
\Gamma _{\ \mu \nu }^{\alpha }\equiv \frac{1}{2}g^{\alpha \lambda }(g_{\mu\lambda ,\nu }+g_{\lambda \nu ,\mu }-g_{\mu \nu ,\lambda }),
\end{equation}
and the \textit{Disformation tensor} $L_{\ \mu \nu }^{\alpha }$\ is defined as:
\begin{equation}\label{eq:3}
L_{\ \mu \nu }^{\alpha }\equiv \frac{1}{2}(Q_{~~\mu\nu }^{\alpha }-Q_{\mu~~\nu }^{\ \alpha }-Q_{\nu~~\mu }^{\ \alpha })=L_{\ \nu \mu }^{\alpha }.
\end{equation}
Finally, the expression for the \textit{Contortion tensor} $K_{\ \mu \nu}^{\alpha }$\ is expressed as:
\begin{equation}\label{eq:4}
K_{\ \mu \nu }^{\alpha }\equiv \frac{1}{2}(T_{~~\mu \nu }^{\alpha }+T_{\mu~~\nu }^{~\alpha }+T_{\nu~~\mu }^{~\alpha })=-K_{\ \nu \mu }^{\alpha }.
\end{equation}
The quantities $Q_{\alpha \mu \nu }$ and $T_{\ \mu \nu }^{\alpha }$ in Eqs. \eqref{eq:3} and \eqref{eq:4}, are the non-metricity tensor and the torsion tensor, respectively, which are given as,
\begin{eqnarray}
Q_{\alpha \mu \nu }\equiv \nabla _{\alpha }g_{\mu \nu }\neq 0,\label{eq:5}\\
T_{\ \mu \nu }^{\alpha }\equiv \Sigma _{\ \mu \nu }^{\alpha }-\Sigma _{\ \nu
\mu }^{\alpha }.\label{eq:6}
\end{eqnarray}
As shown in \cite{Jimenez2018}, the connection is presumed to be torsion and curvature free within the \textit{Symmetric Teleparallel Equivalent to General Relativity} (STEGR). The components of the connection in Eq. \eqref{eq:1} can be rewritten as:

\begin{equation}\label{eq:7}
\Sigma ^{\alpha }\,_{\mu \beta }=\frac{\partial y^{\alpha }}{\partial \xi^{\rho }}\partial _{\mu }\partial _{\beta }\xi ^{\rho }.
\end{equation}
In the above equation, $\xi ^{\alpha }=\xi ^{\alpha }(y^{\mu })$ is an invertible relation and $\frac{\partial y^{\alpha }}{\partial \xi ^{\rho }}$ is the inverse of the corresponding Jacobian \cite{Jimenez2020}. This situation is called a coincident gauge, where there is always a possibility of getting a coordinate system with connections $\Sigma _{\ \mu \nu }^{\alpha }$ equaling zero. Hence, in this choice, the covariant derivative $\nabla _{\alpha }$ reduces to the partial derivative $\partial _{\alpha }$ i.e. $Q_{\alpha \mu \nu }=\partial _{\alpha }g_{\mu \nu }$. Thus, it is clear that the Levi-Civita connection $\Gamma _{\ \mu \nu }^{\alpha }$ can be written in terms of the disformation tensor $L_{\ \mu \nu }^{\alpha }$ as $\Gamma _{\ \mu \nu }^{\alpha }=-L_{\ \mu \nu }^{\alpha }$. 
	
The action that conforms with STEGR is described by,
\begin{equation}\label{eq:8}
S_{STEGR}=\int {\frac{1}{2}}\left({Q}\right) {\sqrt{-g}d^{4}x}+\int {\mathcal{L}_{m}\sqrt{-g}d^{4}x},
\end{equation}
where $g$ being the determinant of the tensor metric $g_{\mu \nu }$ and $\mathcal{L}_{m}$ the matter Lagrangian density. The modified $f(R) $ gravity is a generalization of GR, and the $f(T)$ gravity is a generalization of TEGR. Thus, in the same way, the $f(Q) $ is a generalization of STEGR in which the extended action is given by, 
\begin{equation}\label{eq:9}
S=\int {\frac{1}{2}f(Q)\sqrt{-g}d^{4}x}+\int {\mathcal{L}_{m}\sqrt{-g}d^{4}x.}
\end{equation}
Here $f(Q)$ is an arbitrary function of the non-metricity scalar $Q$, with $f\left( Q\right) =Q$ corresponds to the STEGR \cite{Jimenez2018}. In addition, the non-metricity tensor in Eq. \eqref{eq:5} has the following two independent traces,
\begin{equation}\label{eq:10}
Q_{\alpha }=Q_{\alpha }{}^{\mu }{}_{\mu }\text{ \ \ and \ \ }\tilde{Q}_{\alpha }=Q^{\mu }{}_{\alpha \mu }.
\end{equation}
Further, the non-metricity conjugate known as superpotential tensor is given by,
\begin{widetext}
\begin{equation}\label{eq:11}
P^{\lambda}_{~~\mu \nu} \equiv -\frac{1}{4}Q^{\lambda}_{~~\mu \nu} + \frac{1}{4}\left(Q^{~~\lambda}_{\mu~~\nu} + Q^{~~\lambda}_{\nu~~\mu}\right) + \frac{1}{4}Q^{\lambda}g_{\mu \nu} - \frac{1}{8}\left(2 \tilde{Q}^{\lambda}g_{\mu \nu} + {\delta^{\lambda}_{\mu}Q_{\nu} + \delta^{\lambda}_{\nu}Q_{\mu}} \right).
\end{equation}
The non-metricity scalar can be acquired as, 
\begin{equation}\label{eq:12}
Q=-Q_{\lambda \mu \nu }P^{\lambda \mu \nu }.
\end{equation}
Now, the energy-momentum tensor of the content of the Universe as a perfect fluid matter is given as,
\begin{equation}\label{eq:13}
T_{\mu \nu }=\frac{-2}{\sqrt{-g}}\frac{\delta (\sqrt{-g}\mathcal{L}_{m})}{\delta g^{\mu \nu }}.
\end{equation}
By varying the above action \eqref{eq:9} with regard to the metric tensor $g_{\mu \nu }$\ components yield, 

\begin{equation}\label{eq:14}
\frac{2}{\sqrt{-g}}\nabla_\lambda (\sqrt{-g}f_Q P^\lambda\:_{\mu\nu}) + \frac{1}{2}g_{\mu\nu}f+f_Q(P_{\mu\lambda\beta}Q_\nu\:^{\lambda\beta} - 2Q_{\lambda\beta\mu}P^{\lambda\beta}\:_\nu) = -T_{\mu\nu}.
\end{equation}
\end{widetext}
Here, for simplicity we consider $f_{Q}=\frac{df}{dQ}$. Again, by varying the action with regard to the connection, we can get, 
\begin{equation}\label{eq:15}
\nabla _{\mu}\nabla _{\nu}(\sqrt{-g}f_{Q}P_{~~\mu\nu}^{\lambda} + H_{~~\mu\nu}^{\lambda})=0,
\end{equation}
where $H_{~~\mu\nu}^{\lambda}=-\frac{1}{2}\frac{\delta(\sqrt{-g}\mathcal{L}_{m})}{\delta\Gamma _{~~\mu\nu}^{\lambda}}$ denotes the hyper momentum tensor density. Furthermore, we may deduce the additional restriction over the connection, $\nabla _{\mu}\nabla _{\nu}( H_{~~\mu\nu}^{\lambda})=0$. As follows from Eq. \eqref{eq:15},
\begin{equation*}
\nabla _{\mu}\nabla _{\nu}(\sqrt{-g}f_{Q}P_{~~\mu\nu}^{\lambda} )=0.
\end{equation*}

While considering the variation of action with respect to connection, there are two ways to incorporate symmetric teleparallelism. The first approach involves the use of inertial variation \cite{Golovnev2017}, where the connection is set in its pure-gauge form within the action. The second approach involves the consideration of a general connection in the action, but with the introduction of Lagrange multipliers to compensate for curvature and torsion \cite{Jimenez2018a}.

\section{FLRW Universe in $f(Q)$ cosmology}\label{Sec:III}
On large scale, i.e. a scale greater than that of galaxy clusters, the Universe is homogeneous and isotropic. Here, we consider a flat FLRW space time in the Cartesian coordinate as:
\begin{equation}\label{eq:16}
ds^{2}=-dt^{2}+a^{2}(t)[dx^{2}+dy^{2}+dz^{2}],
\end{equation}
where $a(t)$ is the scale factor of the Universe. The non-metricity scalar corresponding to the metric \eqref{eq:16} can be obtained as $Q=6H^{2}$, where $H=\frac{\dot{a}}{a}$ is the Hubble parameter that measures the rate of expansion of the Universe. An overdot on $a(t)$ represents the derivative in cosmic time. In cosmology, the most commonly used matter component is the perfect cosmic fluid whose energy-momentum tensor is given by:
\begin{equation}\label{eq:17}
T_{\mu\nu}=(\rho+p)u_{\mu}u_{\nu}+pg_{\mu\nu}.
\end{equation}
Here, $\rho $ and $p$\ respectively represent the energy density and isotropic pressure of the perfect fluid, and $u^{\mu }=(1,0,0,0)$ represents the four-velocity vector components characterizing the fluid. So, the modified Friedmann equations that describe the dynamics of the Universe in $f(Q)$ gravity using coincidence gauge are \cite{Jimenez2018}:
\begin{eqnarray}
Qf_{Q}-\frac{f}{2} &=& \rho~,\label{eq:18}\\
f_{Q}\dot{H}+\dot{f_{Q}}H &=& -\frac{1}{2}(\rho+p)~.\label{eq:19} 
\end{eqnarray}
It is worth noting that the standard Friedmann equations of GR can be obtained if $f(Q)=Q$ is substituted \cite{Lazkoz2019}.
	
The standard matter field satisfies the continuity equation $\dot{\rho}=-3H(\rho+p)$ which is consistent with the above cosmological equations \cite{Jimenez2018}. Here, we consider the matter component composed of the pressureless matter and radiation whose conservation equation can be respectively written as: 
\begin{equation}\label{eq:20}
\dot{\rho}_{m}+3H\rho_{m} =0, ~~~~~~~~~~~~~~~~~~~~~~ \dot{\rho}_{r}+4H\rho_{r} = 0.
\end{equation}
From Eq. \eqref{eq:20}, $\rho_{m} \propto a^{-3}$ and $\rho_{r} \propto a^{-4}$. Here, $\rho_{m}, \rho_{r}$ denotes the energy density of matter and radiation, respectively. Using the form $f(Q) = Q+F(Q)$ in Eqs. \eqref{eq:18} and \eqref{eq:19}, one may also the express the effective energy density $\rho $ and effective pressure $p$ of the fluid in the geometrical terms as \cite{Khyllep2021},
\begin{eqnarray}
\rho &=&  \rho_{m}+\rho_{r}+\underbrace{\frac{F-2QF_{Q}}{2}}_{\rho_{de}},\label{eq:21}\\
p &=& p_{r}+\underbrace{QF_{Q}+2\dot{H}(2QF_{QQ}+F_{Q})-\frac{F}{2}}_{p_{de}}.\label{eq:22}
\end{eqnarray}
Note that $p_m=0$ and the terms $\rho_{r}$ and $p_{r}$ are used to analyze the behavior of the model's energy density and pressure. Furthermore, using Eqs. \eqref{eq:21} and \eqref{eq:22}, we can introduce the equation of state (EoS) for dark energy (\(\omega_{de}\)) and an effective EoS (\(\omega\)), which determine the overall accelerated expansion of the Universe. These are given by the following equations:

\begin{widetext}
\begin{eqnarray}
\omega_{de} &=& \frac{p_{de}}{\rho_{de}} =-1+\frac{4\dot{H}(2QF_{QQ}+F_{Q})}{F-2QF_{Q}},\nonumber\\
\omega &=& \frac{p}{\rho} = \frac{-F+4\dot{H} (F_{Q}+2QF_{QQ})+2(QF_{Q}+p_{r})}{F-2QF_{Q}+2\rho_{m}+2\rho_{r}}.\label{eq:23}
\end{eqnarray}
\end{widetext}
The model corresponds to $\Lambda CDM$, when $F(Q)=$ constant i.e., $\omega_{de} = -1$; to quintessence, when $-1 < \omega_{de} \leq- \frac{1}{3}$; and to phantom, when $\omega_{de} < -1$. The numerical value of the EoS parameter has been constrained by several cosmological analyses, such as Supernovae Cosmology Project ( $\omega_{de} = -1.035_{-0.059}^{+0.055}$) \cite{Amanullah2010}; WAMP+CMB ($\omega_{de} = -1.079_{-0.089}^{+0.090}$) \cite{Hinshaw2013}; Planck 2018 ($\omega_{de} = -1.03\pm 0.03$ )\cite{Aghanim2020}.

\section{$H(z)$ Parametrization}\label{Sec:IV}
To study the background evolution of the Universe through various evolutionary phases, we consider a well-motivated form $f(Q)$ given by \cite{Harko2018}: 
\begin{equation}\label{eq:24}
f(Q)= \frac{\alpha Q}{Q_{0}} + \frac{\beta Q_{0}}{Q}~,
\end{equation}
where $Q_{0}=6H_{0}^{2}$ ($H_{0}$ is present Hubble value). This form is useful in describing the late time acceleration without invoking the DE component \cite{Jimenez2020}. However, for $\beta=0$, we have $F=0$, $\omega_{de}$ becomes undefined. The relation between the Hubble parameter and redshift is given by \cite{Narawade2022b}:
\begin{eqnarray}\label{eq:25}
\frac{dH}{dz} &=& \frac{(1+\omega)}{4H(1+z)}\frac{2Qf_{Q}-f}{2Qf_{QQ}+f_{Q}},\nonumber\\
\frac{dH}{dz} &=& \frac{3(1+\omega)}{(1+z)}\frac{H(\alpha H^{4}-3\beta H_{0}^{4})}{2(\alpha H^{4}+3\beta H_{0}^{4})},
\end{eqnarray}
where we use the general density-pressure relation, i.e. $\frac{p}{\rho}=\omega$. Using Eq. \eqref{eq:24} in Eq. \eqref{eq:25}, we obtain:
\begin{equation}\label{eq25.1}
    \alpha H^{4}-DH^{2}(1+z)^{3(1+\omega)}-3\beta H_{0}^{4} = 0,
\end{equation}
where $D$ is the constant of integration. This is quadratic equation in $H^{2}$.\\
Solving Eq. \eqref{eq25.1} for $H^{2}$ and considering positive root for $H^2$, we get \cite{Sahni2003a}
\begin{equation}\label{eq25.2}
    H^{2} = H_{0}^2\left[A(1+z)^{B}+\sqrt{A^{2}(1+z)^{2B}+C}\right]
\end{equation}    
where, $B=3(1+\omega)$, $C = \frac{3\beta}{\alpha}$, and $D = 2\alpha AH_{0}^{2}$.
\subsection*{Cosmographic Parameters}
In this subsection, we introduce some cosmographic parameters, such as the deceleration parameter \( q \) and the state finder parameters \(\{j, s\}\), known respectively as the jerk and snap parameters. These parameters are crucial for distinguishing dark energy models. Cosmographic parameters are geometric quantities that are derived from the scale factor of the Universe, and thus, they depend on the properties of the metric potentials.

 The deceleration parameter \( q(z) \) is defined as:

\begin{equation}\label{eq:26}
q(z) = -1+\frac{(1+z)}{H(z)}\frac{dH(z)}{dz}.
\end{equation}
Eq. \eqref{eq:26} describes the decelerating or accelerating behaviour of the Universe. A positive value of \( q \) indicates a decelerating phase, whereas a negative value corresponds to an accelerating phase. Some cosmological observations have provided its present values as \( q_{0} = -0.51^{+0.09}_{-0.01} \) and \( q_{0} = -0.5422_{-0.0826}^{+0.0718} \). The transition redshift from deceleration to acceleration has been measured as \( z_{t} = 0.65^{+0.19}_{-0.17} \) and \( z_{t} = 0.8596_{-0.2722}^{+2886} \) \cite{Capozziello2008, Capozziello2014, Yang2020}.

Furthermore, the state finder pair is defined as:
 
 \begin{eqnarray}
j(z) &=& q(z) + 2q^{2}(z) + (1+z)\frac{dq(z)}{dz},\label{eq:27} \\
s(z) &=& \frac{j(z)-1}{3\left(q(z)-\frac{1}{2}\right)}, ~~~~~~~~~~~~~~\left( q\neq \frac{1}{2}\right).
\label{eq:28}
\end{eqnarray}

Based on the expressions of these parameters, Sahni et al. \cite{Sahni2003, Alam2003, Zhang2005} categorize the following:
\begin{itemize}
    \item \( (j=1,~s=0) \rightarrow \Lambda \)CDM;
    \item \( (j<1,~s>0) \rightarrow \) Quintessence;
    \item \( (j>1,~s<0) \rightarrow \) Chaplygin Gas;
    \item \( (j=1,~s=1) \rightarrow \) SCDM.
\end{itemize}
It is noteworthy that \( \{j,~s\} = \{1, 0\} \) represents the point corresponding to the flat \( \Lambda \)CDM model. By using this as a reference point, one can assess the deviation of other models from the flat \( \Lambda \)CDM model.

In the \( \{j,~s\} \) plane, positive \( s \) and negative \( s \) respectively indicate quintessence-like and phantom-like dark energy (DE) models respectively. Moreover, when \( \{j,~s\} \) passes through the point \( \{1, 0\} \) in the \( j-s \) plane, it signifies a transition from phantom to quintessence behaviour. Thus, the state finder pair provides a robust method for classifying DE models \cite{Wang2009}.

\section{Observational constraints}\label{Sec:V}
In this section, we will use data sets from various cosmological observations to constrain the free parameters of $H(z)$. Specifically, we will use data that describes the distance-redshift relationship. The datasets we will use include the expansion rate data from early-type galaxies ($H(z)$ data), $Pantheon^{+}$ data, BAO (Baryon Acoustic Oscillations) data, and CMB (Cosmic Microwave Background) distance priors. These observational datasets are independent of any specific cosmological model and serve as tools for estimating cosmological parameters. A brief discussion of each dataset is given below:

\subsection{Monte Carlo Markov Chain (MCMC)}
The best-fit values of the free parameters are constrained by considering the $H(z)$ and $Pantheon^{+}$ data sets using the MCMC process. To perform this, we use the Scipy optimization technique from the Python library, combined with the emcee package, and apply a Gaussian prior with a fixed $\sigma = 1.0$ as the dispersion. The diagonal panels of the MCMC plot show the $1$-D marginalized distribution for each model parameter, with a thick line indicating the best-fit value. The off-diagonal panels display the $2$-D projections of the posterior probability distributions for each pair of parameters, with contours indicating the $1-\sigma$ and $2-\sigma$ confidence regions. 

The main objective of this technique is to maximize the total likelihood function $\mathcal{L}_{\rm tot} \simeq e^{-\chi^2/2}$, which is equivalent to minimizing the total $\chi^2$. Here, $\chi^2$ is determined by the contributions from the $H(z)$ and $Pantheon^{+}$ data sets, denoted as $\chi^{2}_{H(z)}$ and $\chi_{SN}^2$, respectively. This approach allows for a comprehensive and determination of the model parameters, ensuring consistency with observational data.
	
\subsection{$H(z)$ data}

A list of $32$ correlated data points of the Hubble parameter is provided in the redshift range $0.07 \leq z \leq 1.965$ \cite{Moresco2022, Narawade2022b}. By minimizing the Chi-square value, we determine the mean values of the model parameters $H_{0}$, $A$, and $B$. The Chi-square function from the Hubble data is given as, 

 \begin{equation}\label{eq:30}
\chi_{H(z)}^{2}(p_{s}) = \sum_{i=0}^{32}\frac{\left[H_{th}(z_{i}, p_{s}) - H_{obs}(z_{i})\right]^{2}}{\sigma_{H}^{2}(z_{i})},
\end{equation}
where $H_{obs}(z_{i})$ represents the observed Hubble parameter values, $H_{th}(z_{i}, p_{s})$ represents the theoretical Hubble parameter values based on the model parameters, and $\sigma_{H}^{2}(z_{i})$ is the standard deviation.

\subsection{$Pantheon^{+}$ data}
The $Pantheon^{+}$ sample data set consists of $1701$ light curves of $1550$ distinct Type Ia supernovae, ranging in redshift from $z = 0.00122$ to $2.2613$ \cite{Brout2022}. The model parameters are fitted by comparing the observed and theoretical values of the distance moduli. The $\chi^{2}_{SN}$ function from the $Pantheon^{+}$ sample is given by:

\begin{equation}\label{eq:31}
\chi^{2}_{SN}(z, p_{s}) = \sum_{i,j=0}^{1701} \nabla\mu_{i} \left( C_{SN}^{-1}\right) \nabla\mu_{j},
\end{equation}
where $p_{s}$ represents the model parameters and $C_{SN}$ is the covariance matrix \cite{Scolnic2018}. Once a specific cosmological model has been chosen, the predicted distance modulus $\mu$ is defined as:

\begin{equation}\label{eq:32}
\mu(z, p_{s}) = 5 \log_{10}[d_{L}(z,p_{s})] + \mu_{0},
\end{equation}
where $\mu_{0}$ is the nuisance parameter and $d_{L}$ is the dimensionless luminosity distance defined as:
\begin{equation}\label{eq:33}
d_{L}(z) = (1+z)\int_{0}^{z}\frac{d\tilde{z}}{E(\tilde{z})} ,
\end{equation}
with $E(z) = \frac{H(z)}{H_{0}}$ as the dimensionless parameter, and $\tilde{z}$ as the variable of integration from $0$ to $z$.

\subsection{$BAO/CMB$ Constraints}
We verify the obtained values of $H(z)$ parameters presented in Table- \ref{table:I} with the $BAO/CMB$ data. The $\chi^{2}$ for the $\textit{BAO/CMB}$ analysis using acoustic scale is given as
\begin{equation*}
\chi_{\textit{BAO/CMB}}^{2} = X^{T}C^{-1}X,
\end{equation*}
where $X$ depends on the survey considered and the inverse of covariance matrix $C$ is given by \cite{Giostri2012},
\begin{widetext}
\( X= 
	\renewcommand\arraystretch{1.4}
	\begin{pmatrix}
		\frac{d_{A}(z_{*})}{D_{V}(0.106)}-30.95 \\
		\frac{d_{A}(z_{*})}{D_{V}(0.200)}-17.55 \\
		\frac{d_{A}(z_{*})}{D_{V}(0.350)}-10.11 \\
		\frac{d_{A}(z_{*})}{D_{V}(0.440)}-8.44 \\
		\frac{d_{A}(z_{*})}{D_{V}(0.600)}-6.69 \\
		\frac{d_{A}(z_{*})}{D_{V}(0.730)}-5.45
	\end{pmatrix}
	\) ~~and~~
	\(
	C^{-1} = 
	\renewcommand\arraystretch{1.4}
	\begin{pmatrix}
		0.48435 & -0.101383 & -0.164945 & -0.0305703 & -0.097874 & -0.106738 \\
		-0.101383 & 3.2882 & -2.45497 & -0.0787898 & -0.252254 & -0.2751 \\
		-0.164945 & -2.45497 & 9.55916 & -0.128187  & -0.410404 & -0.447574  \\
		-0.0305703 & -0.0787898 & -0.128187 & 2.78728 & -2.75632 & 1.16437 \\
		-0.097874 & -0.252254 & -0.410404 & -2.75632 & 14.9245 & -7.32441 \\
		0.106738 & -0.2751 & -0.447574 & 1.16437 & -7.32441 & 14.5022 \\
	\end{pmatrix}
	\)\,.
\end{widetext}
The expression for the comoving angular-diameter distance [$d_{A}(z_{*})$] and the dilation scale [$D_{V}(z)$] are respectively given by:
\begin{eqnarray}
d_{A}(z_{*}) &=& \int_{0}^{z_{*}}\frac{d\tilde{z}}{H(\tilde{z})}, \nonumber\\
D_{V}(z) &=& \left[\frac{(d_{A}(z))^{2}cz}{H(z)}\right]^{\frac{1}{3}}.\label{eq:34} 	\end{eqnarray}
The epoch at which baryons were released from photons is called drag epoch $(z_{d})$. At this epoch, the photon pressure is no longer able to avoid the gravitational instability of the baryons. The value $z_{d} = 1020$ \cite{Komatsu2009} is commonly used as an approximate reference for the drag epoch or the decoupling of photons and baryons in the early Universe. At $z_{d} = 1020$, the Universe had expanded and cooled significantly. This allows photons and baryons to decouple and photons travels freely.

\section*{Results}
After implementing the MCMC process, we obtained the best-fit values for the cosmological free parameters $H_{0}$, $A$, $B$, and $C$, as shown in Table- \ref{table:I}. Fig. \ref{fig:I} presents the confidence contours along with the marginalized posterior distribution of the different cosmological parameters from the $H(z)$ and $Pantheon^{+}$ data sets. The constrained values for the free parameters $H_{0}$, $A$, $B$, and $C$ using these data sets are summarized in Table- \ref{table:I}. 

Fig. \ref{fig:II} shows the best-fit curves for the model described by Eq. \eqref{eq25.2} compared with the standard $\Lambda$CDM model and various data points. An interesting feature is observed in the value of the Hubble constant, defined as $H_{0} = 100 h$. Our results for $H_{0}$ closely match the value forecasted by Planck $\Lambda$CDM estimations \cite{Aghanim2020}. Both the solid red line and the dashed black line appear within the error bars, indicating a good fit to the data.

\begin{widetext}
  
 \begin{table}[H]
		\fontsize{10pt}{4pt}
		\addtolength{\tabcolsep}{0.5pt}
		\begin{center}
  \renewcommand{\arraystretch}{1.5}
			\begin{tabular}{ccccc}
				\hline
				\toprule
				\hline
				Data sets & $H_{0}$ & $A$ & $B$ & $C$ \\ [0.1cm] 
				\hline\hline\\[-0.1cm]  
				$H(z)$ & $69.49\pm 1.10$ & $0.19\pm 0.99$ & $2.66^{+1.10}_{-0.97}$ & $0.55^{+0.95}_{-1.10}$\\[0.1cm] 
                $Pantheon^{+}$ & $69.5^{+2.10}_{-1.80}$ & $0.20^{+2.00}_{-2.20}$ & $2.70\pm 2.10$ & $0.5^{+2.20}_{-2.10}$\\[0.1cm]
                $H(z)+Pantheon^{+}+BAO$ & $70.16^{+1.00}_{-0.90}$ & $0.19^{+0.15}_{-0.13}$ & $2.73\pm 0.10$ & $0.49^{+0.08}_{-1.00}$\\[0.1cm]
				\hline
			\end{tabular}
		\caption{The marginalized $1-\sigma$ values of free cosmological parameters.}\label{table:I}
		\end{center}
	\end{table}

 \begin{figure}[H]
\centering
\includegraphics[width=8.92cm,height=9cm]{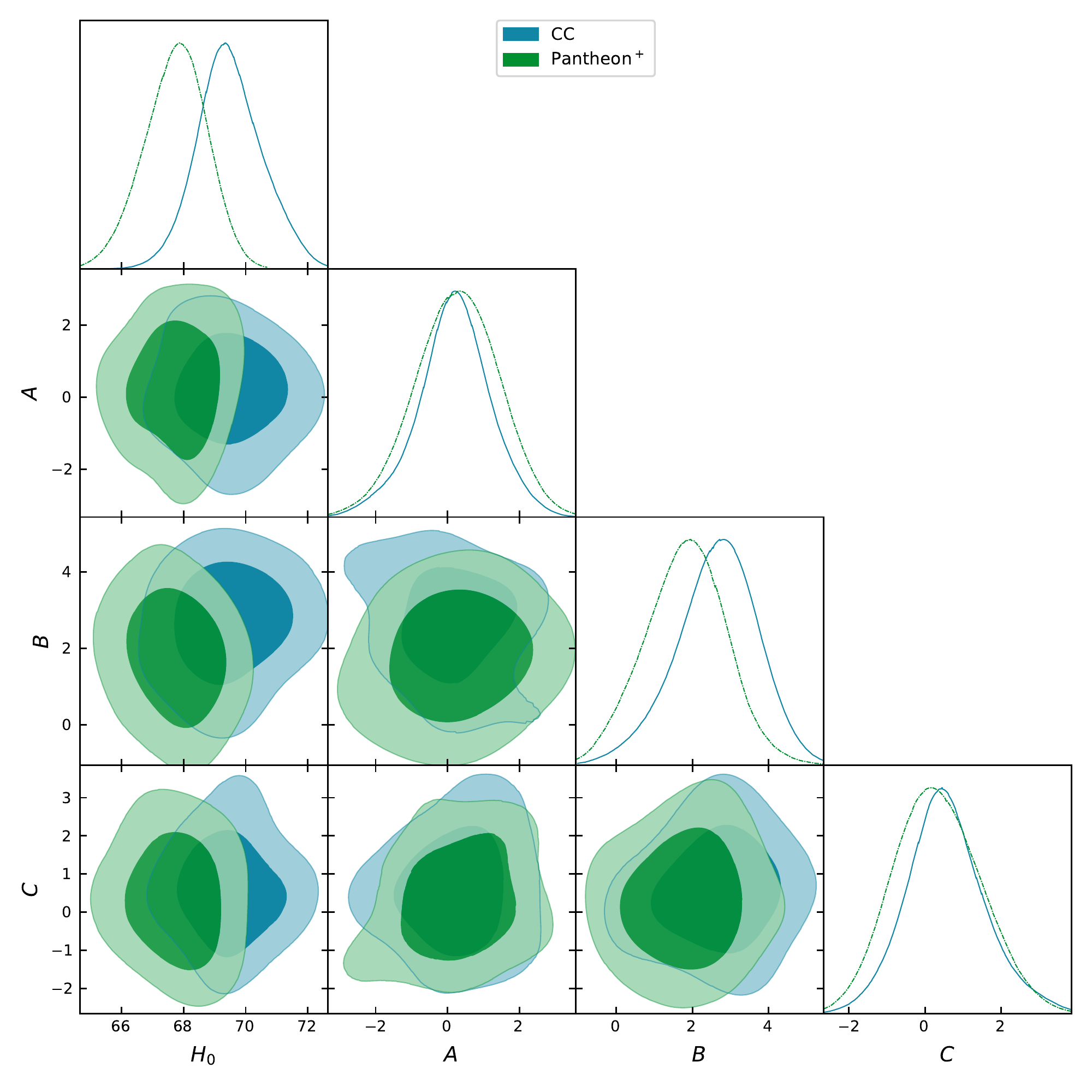}
\includegraphics[width=8.92cm,height=9cm]{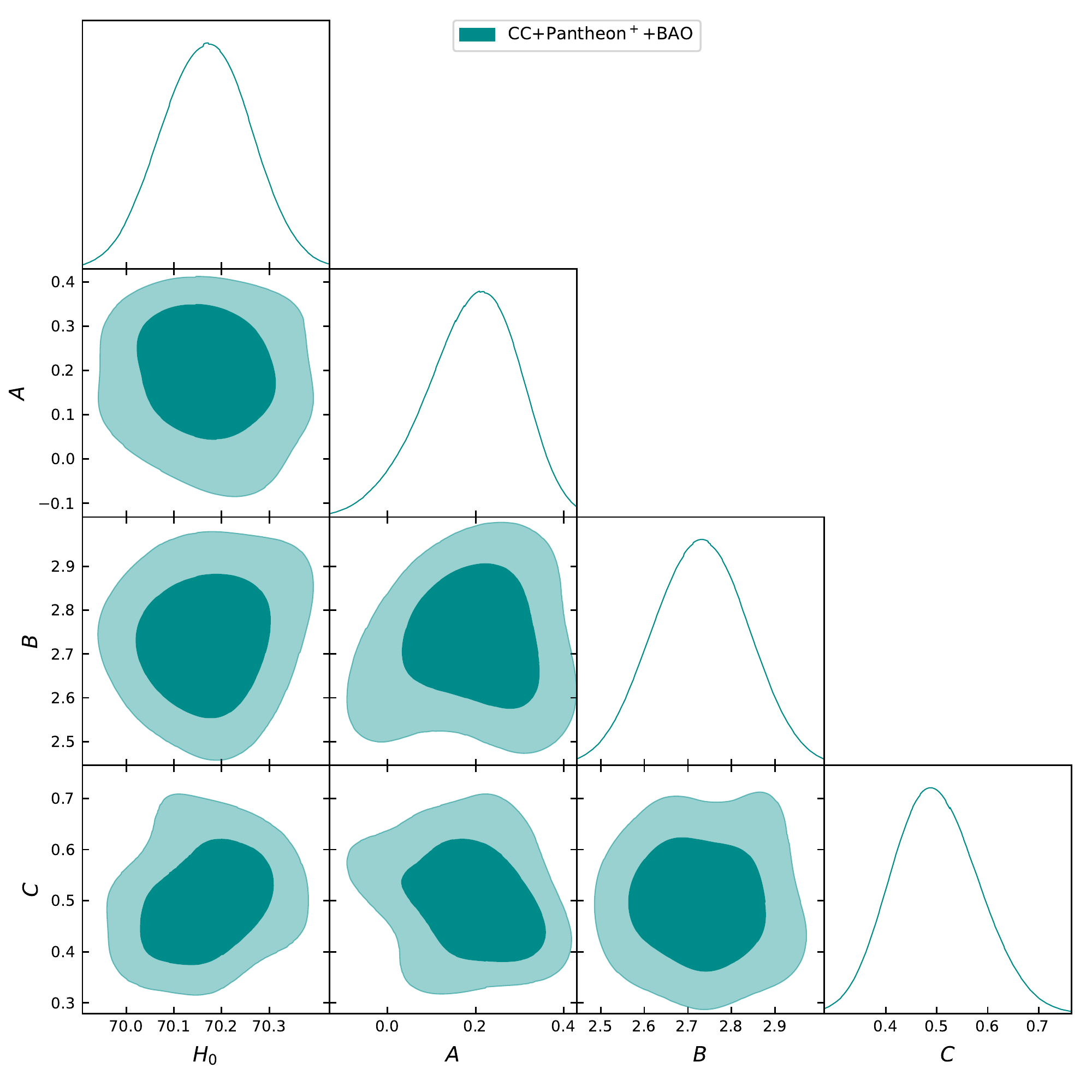}
\caption{MCMC 1$-\sigma$ and 2-$\sigma$ confidence contour plot obtained from $H(z)$ and $Pantheon^{+}$ data set ({\bf Left panel}) and $H(z)+Pantheon^{+}+BAO$ data set ({\bf Right panel}).} \label{fig:I}
\end{figure} 
\begin{figure}[H]
\centering
\includegraphics[scale=0.5]{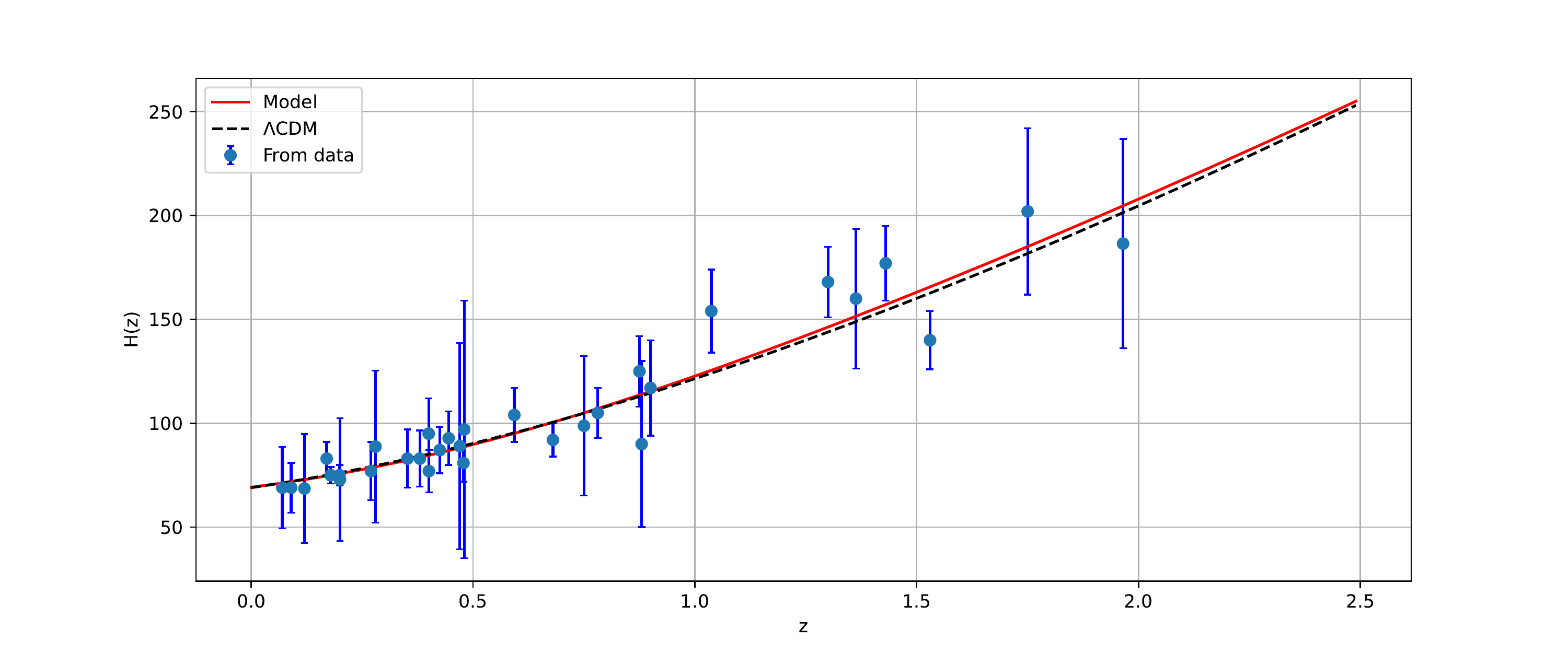}
\includegraphics[scale=0.5]{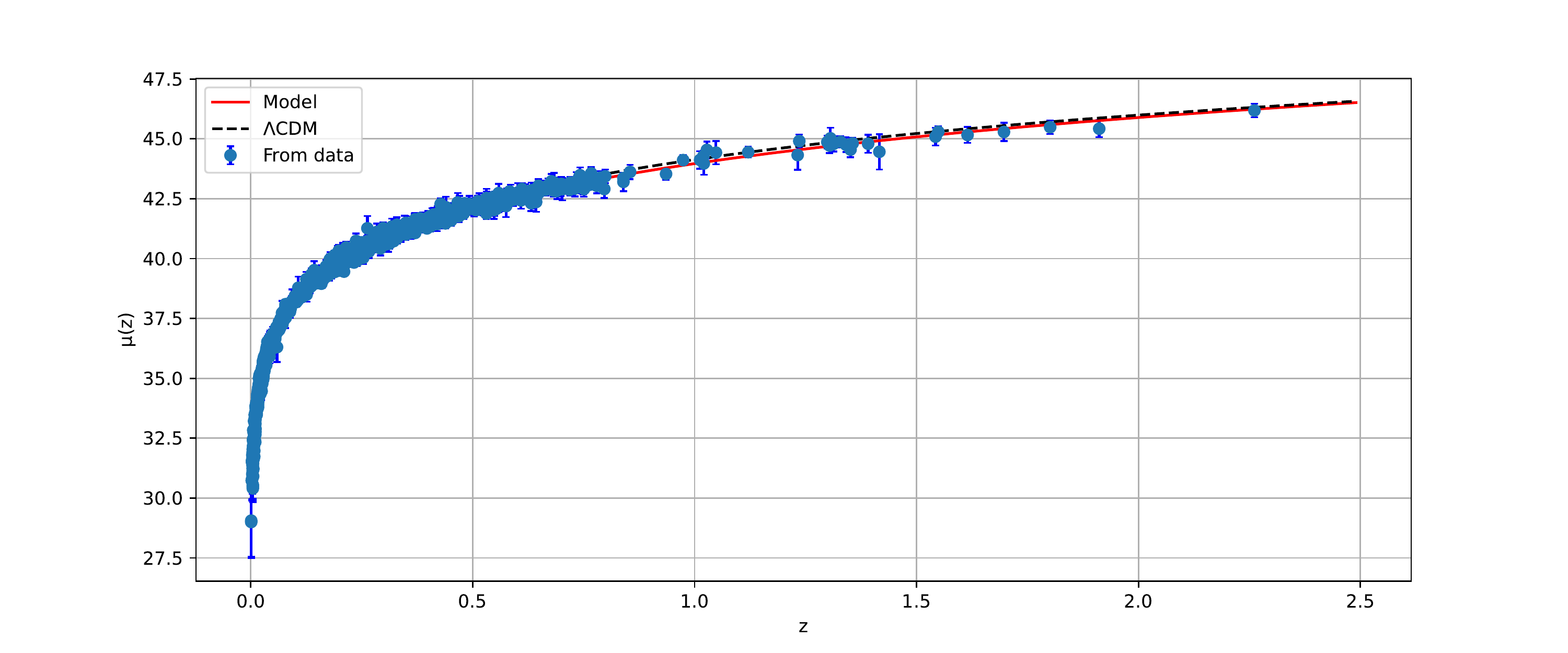}
\caption{ $H(z)$ in redshift from 32 data points of $H(z)$ data ({\bf Upper panel}). $\mu(z)$ in redshift in $Pantheon^{+}$ data set ({\bf Lower panel}). Blue error bars from the data set, red solid line for the model and broken black line for $\Lambda$CDM.}
\label{fig:II}
\end{figure}

\begin{figure}[H]
\centering
\includegraphics[scale=0.5]{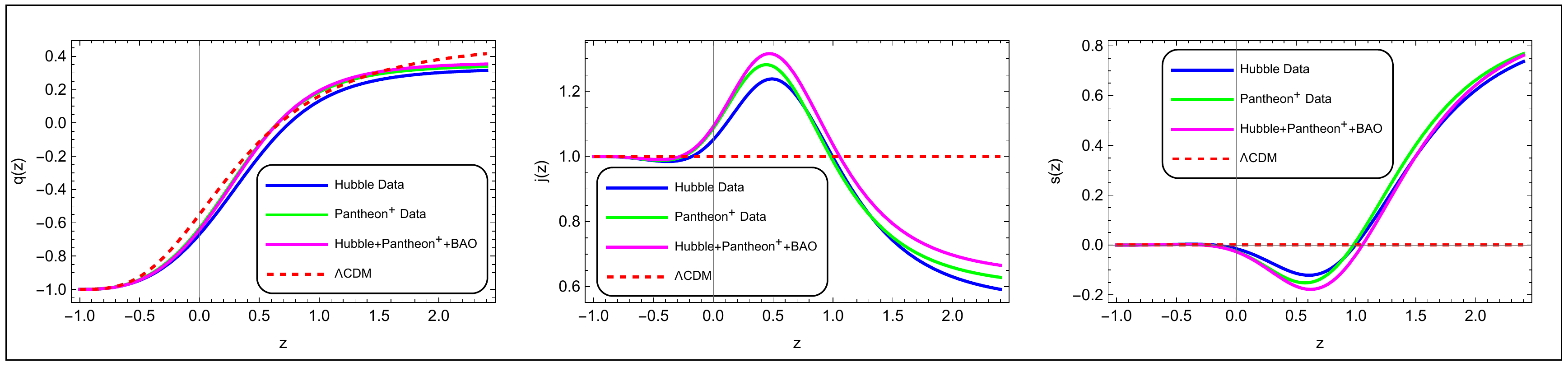}
\caption{Behaviour of deceleration parameter ({\bf Left panel}), jerk parameter ({\bf Middle panel}), snap parameter ({\bf Right panel}) versus redshift.}
\label{fig:III}
\end{figure}
\end{widetext}
Once we obtained the range of the free parameters, we can further examine the behavior of other geometrical parameters (refer to \hyperref[Appendix]{Appendix}). From Fig. \ref{fig:III} (Left panel), we observe that $q(z)$ shows a transition from deceleration to acceleration. The transition points for the $H(z)$ data set, $Pantheon^{+}$ data set, and $H(z)+Pantheon^{+}+BAO$ are $z_{t} = 0.73$, $z_{t} = 0.65$, and $z_{t} = 0.67$, respectively, for the best-fit values obtained in Table- \ref{table:I}. The present values of the deceleration parameter for each data set are summarized in Table- \ref{table:II}.

Fig. \ref{fig:III} (Middle panel) shows the behavior of the jerk parameter. It can be seen that the $H(z)$ data set exhibits less deviation in $j(z)$ than the $Pantheon^{+}$ data set, which in turn shows less deviation than the $H(z)+Pantheon^{+}+BAO$ data set for the value $j(z)=1$. Similarly, the snap parameter for the $H(z)+Pantheon^{+}+BAO$ data set shows more deviation than the snap parameter for the $Pantheon^{+}$ data set, and the $Pantheon^{+}$ data set shows more deviation than the $H(z)$ data set for the value $s(z)=0$ (Fig. \ref{fig:III}, Right panel). The $(j,~s)$ values for the $H(z)$, $Pantheon^{+}$, and $H(z)+Pantheon^{+}+BAO$ data sets are $(1.05, -0.01)$, $(1.08, -0.02)$, and $(1.09, -0.03)$, respectively.

\begin{table}
		\fontsize{10pt}{4pt}
		\addtolength{\tabcolsep}{0.5pt}
		\begin{center}
  \renewcommand{\arraystretch}{1.5}
			\begin{tabular}{cccc}
				\hline
				\toprule
				\hline
				Data sets & $q_{0}$ & $j_{0}$ & $s_{0}$\\[0.1cm]
				\hline\hline\\[-0.1cm]  
				$H(z)$ & $-0.66$ & $1.05$ & $-0.01$ \\[0.1cm] 
                $Pantheon^{+}$ & $-0.63$ & $1.08$ & $-0.02$ \\[0.1cm]
                $H(z)+Pantheon^{+}+BAO$ & $-0.64$ & $1.09$ & $-0.03$ \\[0.1cm]
				\hline
			\end{tabular}
		\caption{Present value of deceleration, jerk and snap parameter based on $1-\sigma$ values of free parameters.}\label{table:II}
		\end{center}
	\end{table}

\section{State-finder and $Om(z)$ Diagnostics}\label{Sec:VI}

It is possible to validate any cosmological model through both theoretical and observational tests. In this discussion, we will explore state-finder and $Om(z)$ diagnostic cosmological tests that may be utilized to validate the model we have derived.

\noindent\textbf{\textit{State-finder Diagnostic:}} The state-finder pair $(j,s)$ characterizes the properties of dark energy in a model-independent manner \cite{Sahni2003, Alam2003}. Therefore, the state-finder pair helps to distinguish between different dark energy models.

\begin{figure}[H]
    \centering
    \includegraphics[width=8cm,height=8cm]{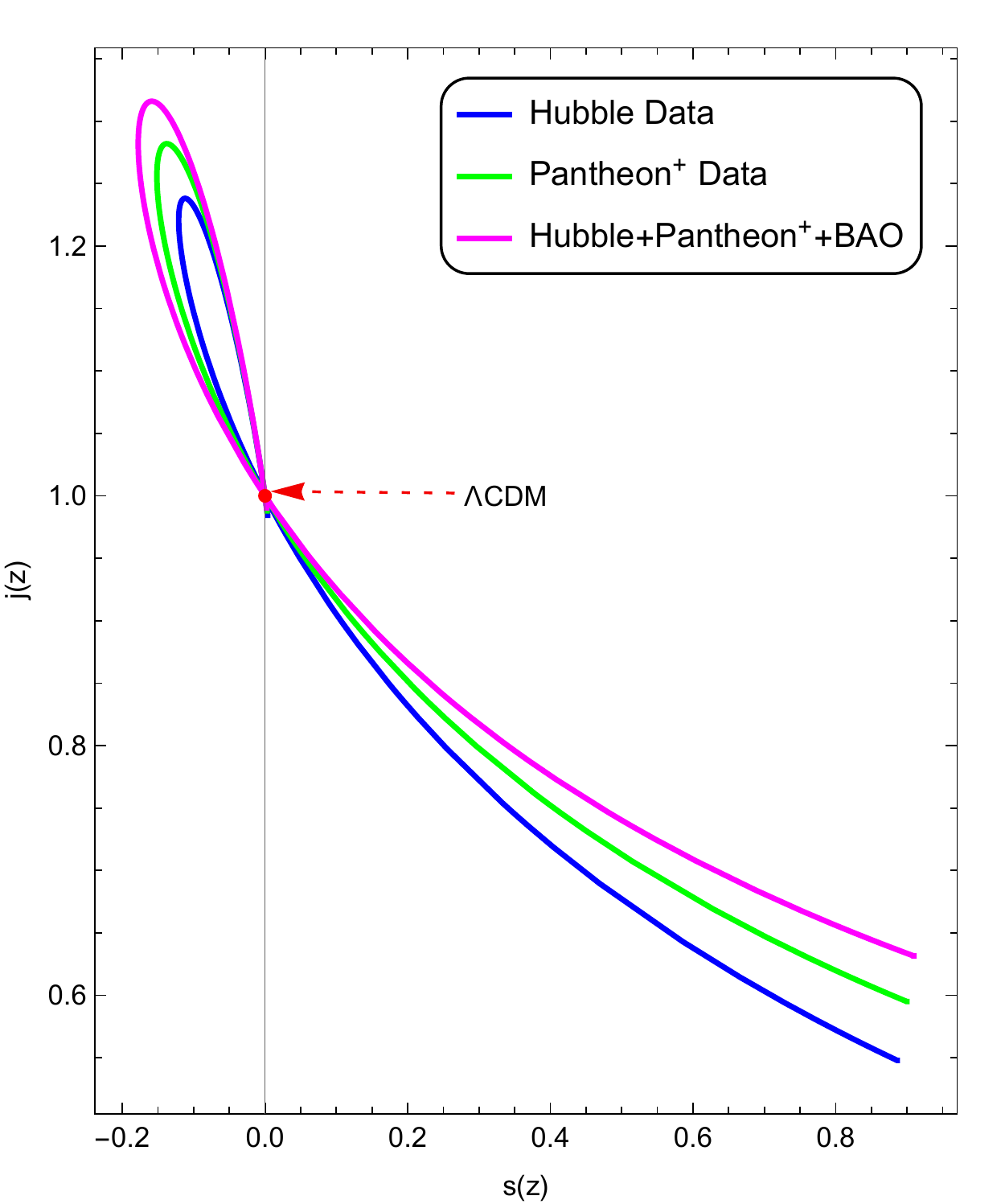}
    \caption{The parametric plot for jerk parameter with respect to snap parameter} 
    \label{statefinder}
\end{figure}
\begin{figure}[H]
    \centering
    \includegraphics[width=8cm,height=8cm]{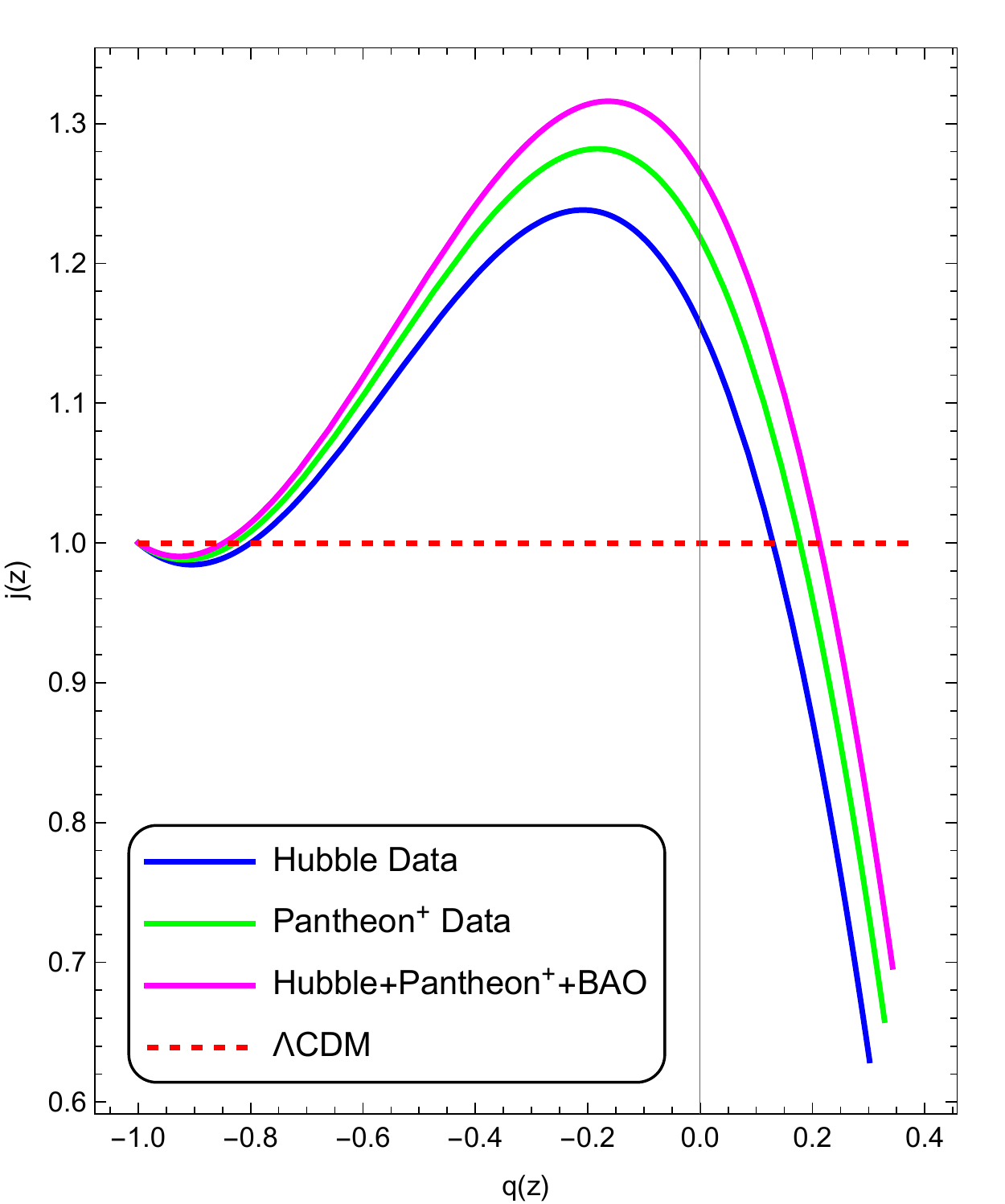}
    \caption{The parametric plot for jerk parameter with respect to deceleration parameter}
    \label{statefinder1}
\end{figure}
\begin{figure}[H]
    \centering
    \includegraphics[width=8cm,height=8cm]{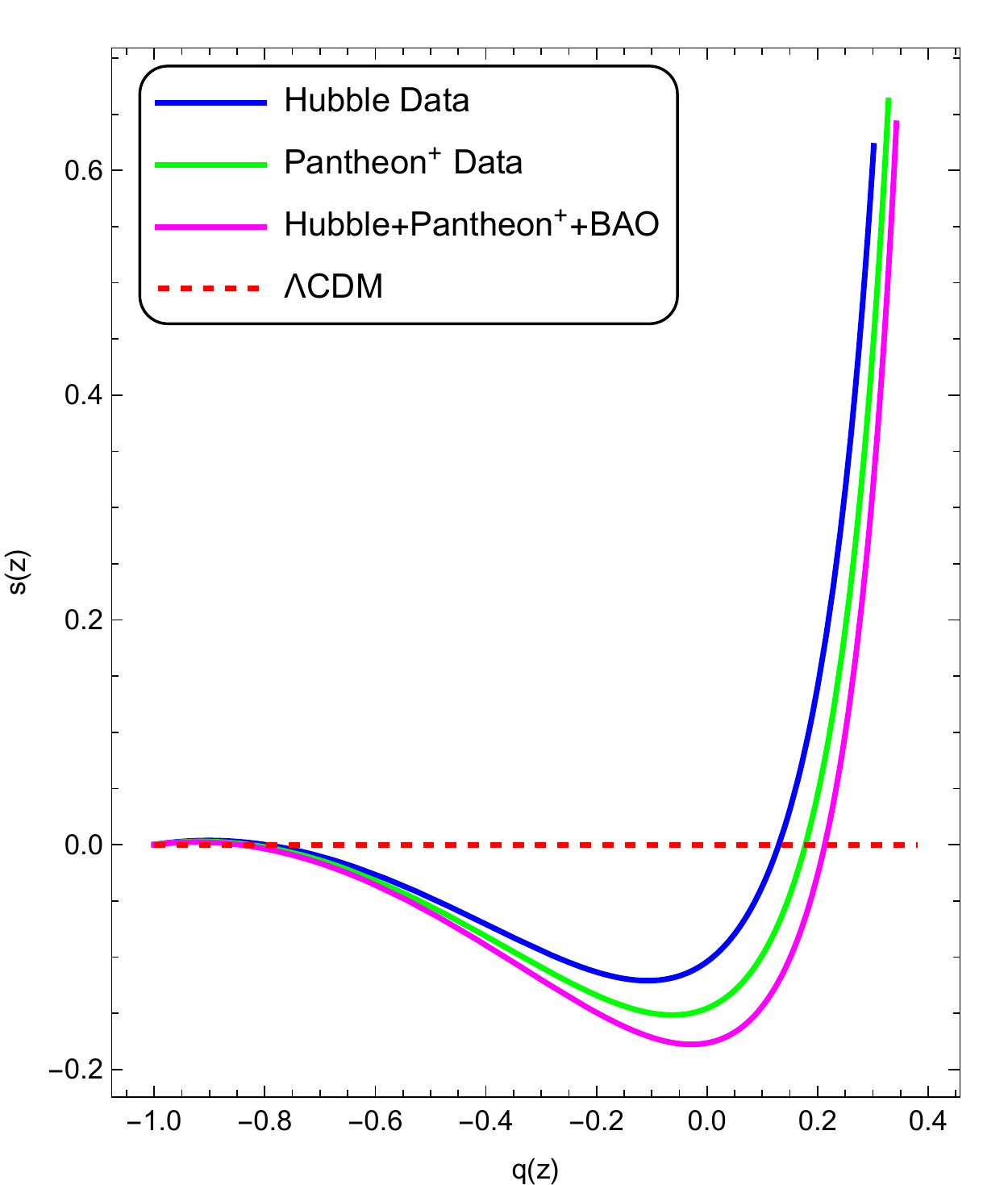}
    \caption{The parametric plot for snap parameter with respect to deceleration parameter }
    \label{statefinder2}
\end{figure}

\noindent In Fig. \ref{statefinder}, one can observe the behavior of the state finder pair, which can be used to determine different dark energy models for the expanding Universe. There are two diagnostic parameters in this method: the $j$ parameter indicates the rate of change in acceleration or deceleration of the Universe, and the $s$ parameter indicates the differences between dark energy models. The plot illustrates the dynamics of the model using constrained parameter values from the dataset. Initially, the model exhibits Quintessence behavior with $(j<1)$ for $(s>0)$. Over time, the $j-s$ pair converges to the $\Lambda$CDM model at the fixed point where $(s=0)$ and $(j=1)$, indicating the late-time accelerating expansion of the Universe. Additionally, the functions $j=j(q)$ [Fig. \ref{statefinder1}] and $s=s(q)$ [Fig. \ref{statefinder2}] demonstrate similar behavior, supporting this interpretation.

\noindent\textbf{\textit{$Om(z)$ Diagnostic:}} The $Om(z)$ diagnostic has been introduced as an alternative approach to test the accelerated expansion of the Universe under the phenomenological assumption that the equation of state (EoS) is $p=\rho\omega$, where $\omega$ is the EoS parameter for a perfect fluid filling the Universe. This diagnostic tool is versatile and can distinguish a wide range of dark energy models, including quintessence, phantom, and $\Lambda$CDM models. The $Om(z)$ diagnostic is particularly sensitive to the EoS parameter, as evidenced by various studies in the literature \cite{Ding2015, Zheng2016, Qi2018}. The nature of the $Om(z)$ slope varies between different dark energy models: a positive slope indicates a phantom phase with $\omega < -1$, while a negative slope indicates a quintessence region with $\omega > -1$. This diagnostic provides a clear method to differentiate between these models, provides understanding into the underlying dynamics of the accelerated expansion of the Universe. The $Om(z)$ diagnostic can be defined as,

\begin{equation*}
Om(z) = \frac{E^{2}(z) - 1}{(1+z)^{3}-1}, \hspace{0.5cm}  E(z)= \frac{H(z)}{H_{0}}
\end{equation*}
\begin{figure}[H]
    \centering
    \includegraphics[width=8cm,height=8cm]{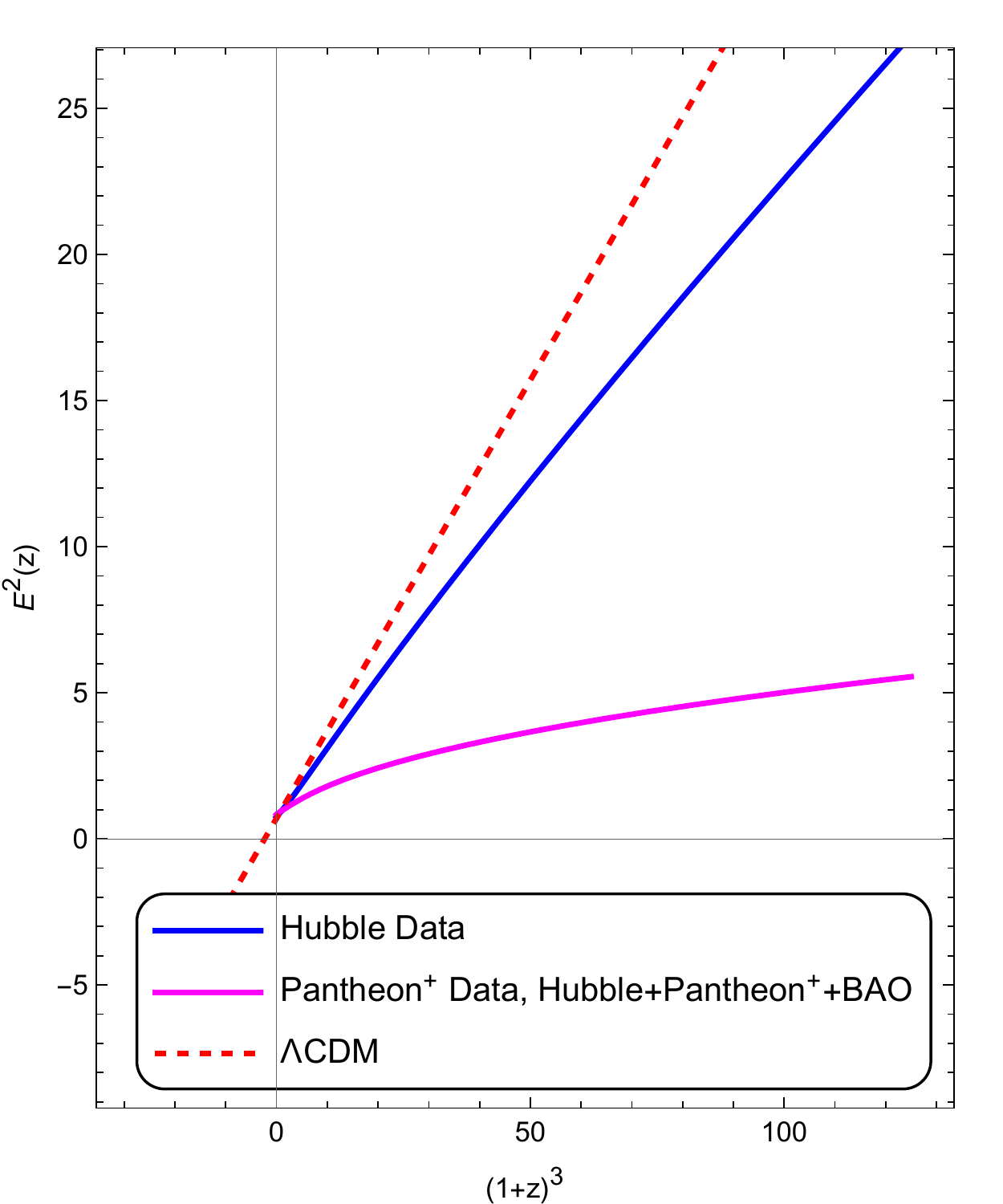}
    \caption{Behaviour of $Om(z)$ in redshift.}
    \label{Om}
\end{figure}
In other words, the $Om(z)$ diagnostic serves as a null test for the cosmological constant. This is because plotting $H^2$ against $(1+z)^3$ results in a straight line for the $\Lambda$CDM model, as illustrated in Fig \ref{Om}. In contrast, for other dark energy models, the $Om(z)$ line is curved. This happens because the relationship $\frac{dH^2}{d(1+z)^3} = \text{constant}$ holds true for quintessence or phantom models only at redshifts significantly greater than one, where the influence of dark energy on the expansion rate can be ignored. The decreasing behavior of $Om(z)$ can be clearly seen in Fig. \ref{Om}, illustrating that the present-time Universe is in a phantom phase and converges to $\Lambda$CDM at late times \cite{Sahni2008}.

\section{ANALYSIS OF THE MODEL}\label{Sec:VII}

In the previous section, the behavior of the geometrical parameters was obtained with the free parameters constrained from various cosmological data sets. Now, we will analyze the dynamical behavior of the Universe. Using Eq. \eqref{eq25.2}, Eqs. \eqref{eq:21}-\eqref{eq:23} can be written as:

\begin{widetext}
\begin{eqnarray}
p &=& \frac{\alpha \left(3H(z)^6+2H(z)^4\frac{dH}{dz}\right)+H_{0}^2 \left(-9\beta H(z)^{2}H_{0}^2+6\beta H_{0}^2\frac{dH}{dz}-18 H(z)^6+2H(z)^4 \left(\Omega_{r}(1+z)^4-6\frac{dH}{dz}\right)\right)}{6H(z)^{4}H_{0}^2},\nonumber\\
\label{eq:40}\\
\rho &=& H(z)^2 \left(3-\frac{\alpha}{2H_{0}^2}\right)+\frac{3\beta H_{0}^2}{2 H(z)^2}+(1+z)^3 (\Omega_{m}+\Omega_{r}z+\Omega_{r}),\nonumber\\
\label{eq:41}\\
\omega &=& \frac{\alpha\left(3H(z)^6+2H(z)^4\frac{dH}{dz}\right)+H_{0}^2 \left(-9\beta H(z)^2 H_{0}^2+6\beta H_{0}^2\frac{dH}{dz}-18 H(z)^6+2H(z)^4 \left(\Omega_{r}(z+1)^4-6\frac{dH}{dz}\right)\right)}{6H(z)^4 H_{0}^2 \left(H(z)^2 \left(3-\frac{\alpha}{2H_{0}^2}\right)+\frac{3\beta H_{0}^2}{2H(z)^2}+(z+1)^3 (\Omega_{m}+\Omega_{r}z+\Omega_{r})\right)}.\nonumber \\\label{eq:42}
\end{eqnarray}

\begin{figure}[H]
\centerline{\includegraphics[scale=0.50]{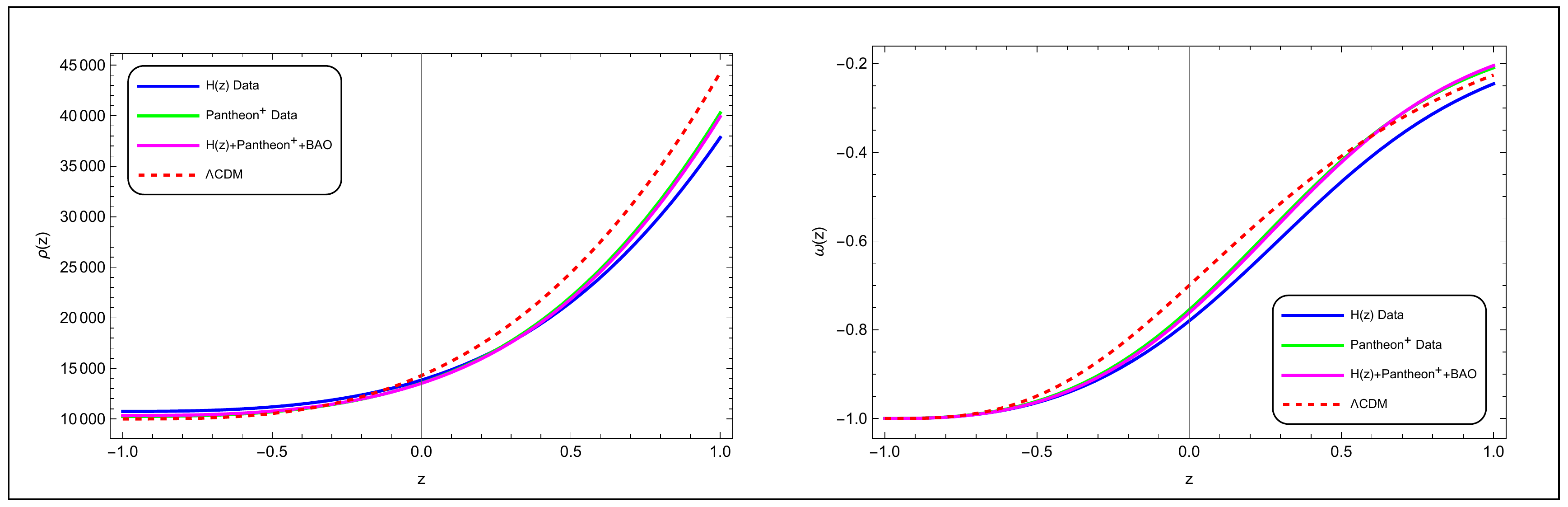}}
\caption{Energy density (\textbf{Left panel}) and EoS parameter (\textbf{Right panel}) in redshift. The parameter scheme: $\Omega_{m}=0.3$ and $\Omega_{r}=0.00001$.}\label{fig:VII}
\end{figure}
\end{widetext}
\newpage
From Fig.~\ref{fig:VII}, we can observe that the effect of the radiation term $\rho_{r}$ reflects here and also the energy density decreases from early times to late times but does not vanish. The total EoS parameter yields present values of $\omega_{0} \simeq -0.78$, $\omega_{0} \simeq -0.77$, and $\omega_{0} \simeq -0.76$ for the constrained free parameters from the $H(z)$ data set, the Pantheon$^{+}$ data set, and the combined $H(z)+Pantheon^{+}+BAO$ data set, respectively. The energy density and EoS parameter are solely dependent on the values of the model coefficients, which govern the evolutionary behavior of the parameters. This model shows the quintessence behavior at present time \cite{Koussour2022b}.

\begin{widetext}

\begin{figure}[H]
 \centerline{\includegraphics[scale=0.50]{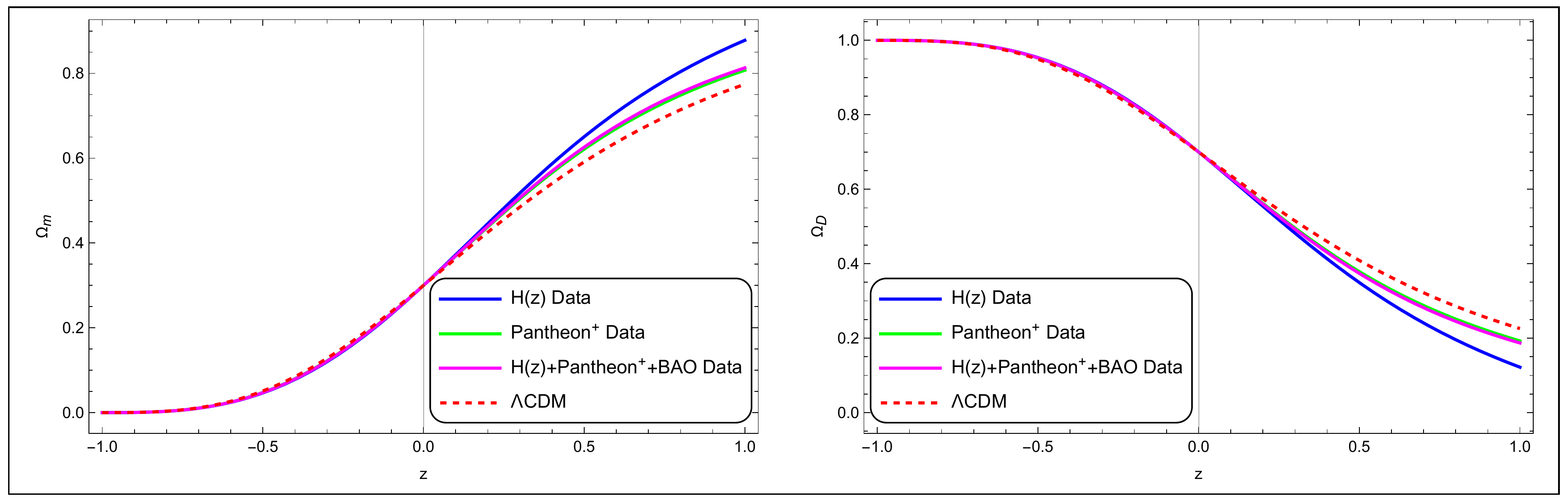}}
\caption{Evolution of Density parameter for matter (\textbf{Left panel}) and dark energy (\textbf{Right panel}) in redshift.}\label{fig:DP}
\end{figure}
\end{widetext}
From the evolution plot of density parameter for matter and dark energy (Fig. \ref{fig:DP}), we can obtain the present value as $\Omega_m\simeq 0.30$ and $\Omega_{de}\simeq 0.70$. The density parameter value for the radiation has been considered to be negligible.

The study of energy conditions is a significant aspect in modified gravity-based cosmological models. These conditions help to ensure the physical viability of such models by imposing constraints on the stress-energy tensor. The general expressions for the energy conditions \cite{Novello2008} are as follows:

\begin{itemize}
\item For each null vector, $T_{ij}u^{i}u^{j} \geq 0 \Rightarrow \rho + p \geq 0$ (Null Energy Condition, NEC).
\item For every time-like vector, $T_{ij}u^{i}u^{j} \geq 0 \Rightarrow \rho \geq 0~~and~~ \rho + p \geq 0$ (Weak Energy Condition, WEC).
\item For any time-like vector, $\left(T_{ij}-\frac{1}{2}Tg_{ij}\right)u^{i}u^{j} \geq 0 \Rightarrow  \rho + 3p \geq 0$ (Strong Energy Condition, SEC).
\item For any time-like vector, $T_{ij}u^{i}u^{j} \geq 0 \Rightarrow \rho - p \geq 0~~and~~ T_{ij}u^{j}$~\text{not~spacelike} (Dominant Energy Condition, DEC).
\end{itemize}

\begin{widetext}

\begin{figure}
\centering
\includegraphics[scale=0.5]{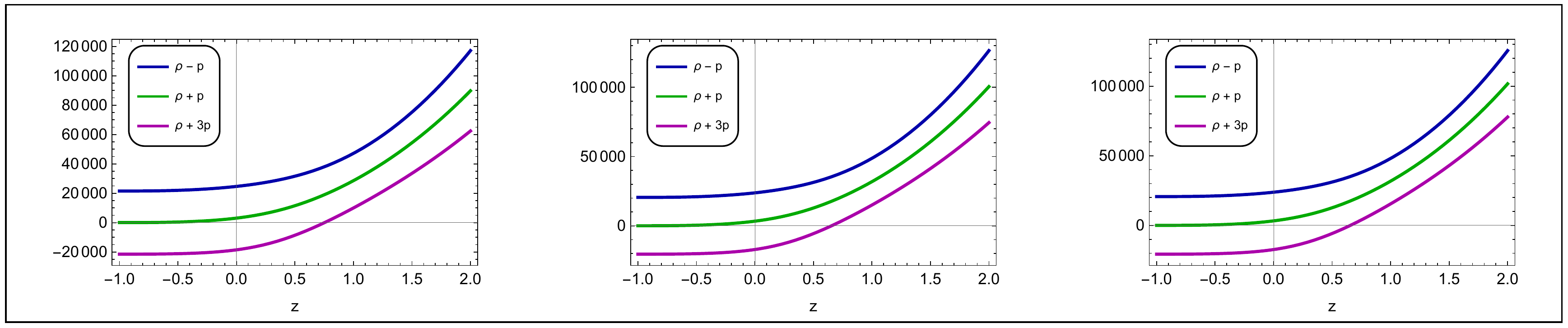}
\caption{Energy conditions in redshift for $H(z)$ data (\textbf{Left panel}), $Pantheon^{+}$ data (\textbf{Middle panel}) and $H(z)+Pantheon^{+}+BAO$ data (\textbf{Right panel}).}
    \label{fig:IX}
\end{figure}
\end{widetext}

By examining these conditions, one can determine the nature of matter and energy in the Universe, which is essential for understanding its accelerated expansion and the role of dark energy.
The energy conditions are basically the boundary conditions that shapes the cosmic evolution \cite{Carroll2003}. Also because of the fundamental casual structure of space time, the gravitational attraction is characterized by the energy conditions \cite{Capozziello2019}. The expressions of energy conditions for the $f(Q)$ gravity model are given in \hyperref[Appendix]{Appendix}. The evolutionary behaviour of the energy conditions is depicted in Fig. \ref{fig:IX}, for the constrained values of the free parameters obtained from the $H(z)$, Pantheon$^+$, and $H(z)+Pantheon^{+}+BAO$ data sets. We observed that during the early epoch, the NEC decreases and remains positive throughout but vanishes at the late epoch. The DEC remains positive throughout and does not violate. While the SEC does not violate at early times, it does violate at late times. These observations highlight the dynamic nature of the energy conditions across different cosmological epochs.

\section{Conclusions}\label{Sec:VIII}
In this work, we have presented a cosmological model within the framework of symmetric teleparallel gravity, specifically focusing on $f(Q)$ gravity. By imposing certain algebraic conditions, we derived the parametric form of the Hubble parameter using a well-motivated form of $f(Q)$. To constrain the free parameters in the Hubble parameterization, we utilized two independent data sets: $H(z)$ and Pantheon$^+$. Additionally, we incorporated data from BAO and CMB distance priors. Our primary objective was to investigate the deviations of $f(Q)$ gravity from the standard $\Lambda$CDM scenario, exploring the potential of this modified gravity theory to provide a viable alternative to the fiducial model.

To achieve this, we employed the MCMC method in conjunction with the Scipy optimization technique from the Python library. Through the successful implementation of the MCMC process, we were able to extract the best-fit values for the cosmological free parameters using the $H(z)$ data set, $Pantheon^+$ data set, and the $BAO/CMB$ data set. Furthermore, for each data set, we analyzed the behavior of the deceleration parameter, indicating the accelerating nature of the Universe, as well as the $(j,s)$ values, which place the Universe within the phantom region. The present values of the cosmographic parameters, obtained using the constrained free parameters, are listed in Table- \ref{table:II}.

We also examined the behaviour of the snap parameter versus the deceleration parameter, as well as the jerk parameter versus the deceleration parameter. These analyses indicate that the Universe converges to $\Lambda$CDM in the late epoch. Additionally, the $Om(z)$ diagnostics further confirm this convergence.
	
Subsequently, we utilized the best-fit parameters to analyze the dynamical behaviour of $f(Q)$ gravity models, focusing on the energy density and the equation of state (EoS) parameter $\omega$. Our analysis revealed that the cosmic energy density is non-negative and decreases from the early Universe to the late Universe. The evolution of $\omega$ demonstrates a transition from a decelerated phase to an accelerated phase. This behaviour exhibited by the $f(Q)$ models effectively describes structure formation and the dominance of dark energy (DE) at the present time. Additionally, at late times, the EoS parameter $\omega$ approaches the limit of a cosmological constant, i.e., $\omega=-1$. Notably, the behaviour of the dark energy EoS parameter $\omega_{de}$ is similar to that of $\omega$ in both data sets, indicating that the dark energy EoS dominates the evolutionary aspects of the models.

Furthermore, we analyzed the energy conditions of the model using both data sets. The null energy condition exhibits a decreasing behaviour and vanishes at late times, while the dominant energy condition remains positive throughout the evolution. The strong energy condition is satisfied in the early Universe but violated at late times for the constrained values of the model parameters derived from the $H(z)$, $Pantheon^+$, and $H(z)+Pantheon^++BAO$ data sets.
 
We wish to mention here that in some of our previous work \cite{Narawade2022b}, we have constrained the free parameters using cosmological data sets by obtaining the exact solution for the Hubble parameter. In the present work, we have used the positive square root for the quadratic equation in $H^2$ to derive the Hubble parameter, and after parameterization, the marginalized values of the free parameters are obtained. It has been observed that the phantom model was obtained in \cite{Narawade2022b}, whereas the present study provides the quintessence behaviour. However, at the later stage of evolution, the model exhibits $\Lambda$CDM behaviour.

In summary, our study successfully demonstrates the viability of $f(Q)$ gravity models as alternatives to the standard $\Lambda$CDM model by analyzing various cosmological parameters and energy conditions. Beyond these, other diagnostic tools like the state finder and $Om(z)$ diagnostics have been used to validate the models further. In future, we will focus on extending the analysis to include different functional forms of $f(Q)$, examining their compatibility with a broader range of observational data, and investigating the potential implications for the early Universe and inflationary scenarios.

\section*{Acknowledgement} BM \& JD acknowledges the support of IUCAA, Pune (India) through the visiting associateship program.
JD was supported by the Core Research Grant of SERB, Department of Science and Technology India (File No. CRG $\slash 2018 \slash 001035$). The authors are thankful to the esteem reviewer for the constructive comments and suggestions to improve the quality of the manuscript.
\newpage
 \begin{widetext}
\section*{Appendix: Expressions of Cosmographic Parameters and Specific Combinations of Energy Density and Pressure}\label{Appendix}
\begin{boxB}
\begin{eqnarray*}
    q(z) &=& -1+\frac{AB(1+z)^B}{2\sqrt{A^2(1+z)^{2B}+C}},\\
    j(z) &=& \frac{-3 A^3 B (1+z)^{3 B}+A^2 \left(B^2+2\right) (1+z)^{2 B} \sqrt{A^2 (1+z)^{2 B}+C}+2 C \sqrt{A^2 (1+z)^{2 B}+C}+A (B-3) B C (1+z)^B}{2 \left(A^2 (1+z)^{2 B}+C\right)^{3/2}},\\
    s(z) &=& -\frac{A B (1+z)^B \left(A (1+z)^B \left(3 A (1+z)^B-B \sqrt{A^2 (1+z)^{2 B}+C}\right)-(B-3) C\right)}{3 \left(A^2 (1+z)^{2 B}+C\right) \left(A B (1+z)^B-3 \sqrt{A^2 (1+z)^{2 B}+C}\right)},
\end{eqnarray*}
 \begin{eqnarray*}
     \rho+p &=& \frac{1}{3} \left(\frac{dH}{dz} \left(\frac{\alpha}{H_{0}^2}+\frac{3\beta H_{0}^2}{H(z)^4}-6\right)+(1+z)^3 (3\Omega_{m}+4\Omega_{r}(1+z))\right)~,\\
    \rho - p &=& \frac{1}{3} \left(-\frac{\alpha\left(3H(z)^2+\frac{dH}{dz}\right)}{H_{0}^2}-\frac{3\beta H_{0}^2 \frac{dH}{dz}}{H(z)^4}+\frac{9\beta H_{0}^2}{H(z)^2}+18H(z)^2+6\frac{dH}{dz}+(1+z)^3 (3 \Omega_{m}+2\Omega_{r}(1+z))\right)~,\\
    \rho + 3p &=& \frac{\alpha \left(H(z)^2+\frac{dH}{dz}\right)}{H_{0}^2}+\frac{3\beta H_{0}^2\frac{dH}{dz}}{H(z)^4}-\frac{3\beta H_{0}^2}{H(z)^2}-6H(z)^2-6\frac{dH}{dz}+(1+z)^3 (\Omega_{m}+2\Omega_{r}(1+z))~.
\end{eqnarray*}
\end{boxB}
\end{widetext}


\begin{thebibliography}{99}
\section*{References}
 	
	\bibitem{Riess1998} A.G. Riess et al., \href{https://doi.org/10.1086/300499}{\textit{The Astronomical Journal}, \textbf{116}, 1009 (1998).}
	
	\bibitem{Perlmutter1999} S. Perlmutter et al., \href{https://doi.org/10.1086/307221}{ \textit{The Astronomical Journal}, \textbf{517}, 565 (1999).}
 
   \bibitem{Komatsu2003} E. Komatsu et al., \href{https://doi.org/10.1086/377226}{\textit{The Astrophysical Journal Supplement Series}, \textbf{148}, 119 (2003).}
		
	\bibitem{Komatsu2011} E. Komatsu et al., \href{https://doi.org/10.1088/0067-0049/192/2/18}{\textit{The Astrophysical Journal Supplement Series}, \textbf{192}, 18 (2011).}
	
	\bibitem{Hinshaw2013} G. Hinshaw, et al., \href{https://doi.org/10.1088/0067-0049/208/2/19}{\textit{The Astrophysical Journal Supplement Series}, \textbf{208}, 19 (2013).}
	
	\bibitem{Bennett2013} C.L. Bennett et al., \href{https://doi.org/10.1088/0067-0049/208/2/20}{\textit{The Astrophysical Journal Supplement Series}, \textbf{208}, 20 (2013).}
	
	\bibitem{Eisenstein2005} D.J. Eisenstein et al., \href{https://doi.org/10.1086/466512}{\textit{The Astrophysical Journal}, \textbf{633}, 560 (2005).}
	
	\bibitem{Daniel2008} S.F. Daniel et al., \href{https://doi.org/10.1103/PhysRevD.77.103513}{\textit{Physical Review D}, \textbf{77}, 105513 (2008).}

       \bibitem{Fowler2010} J.W. Fowler et al., \href{https://doi.org/10.1088/0004-637X/722/2/1148}{\textit{The Astrophysical Journal}, \textbf{722}, 1148 (2010).}
       
        \bibitem{Das2011} S. Das et al., \href{https://doi.org/10.1088/0004-637X/729/1/62}{\textit{The Astrophysical Journal}, \textbf{729}, 62 (2011).}

        \bibitem{Keisler2011} R. Keisler et al., \href{https://doi.org/10.1088/0004-637X/743/1/28}{\textit{The Astrophysical Journal}, \textbf{743}, 28 (2011).}

        \bibitem{Reichardt2012} C.L. Reichardt et al., \href{https://doi.org/10.1088/0004-637X/755/1/70}{\textit{The Astrophysical Journal}, \textbf{755}, 70 (2012).}

	\bibitem{Alam2016} S. Alam, et al., \href{https://doi.org/10.1093/mnras/stx721}{\textit{Monthly Notices of the Royal Astronomical Society}, \textbf{470}, 2617 (2016).}
	
	\bibitem{Ade2016} P.A.R. Ade, et al., \href{https://doi.org/10.1051/0004-6361/201525830}{\textit{Astronomy \& Astrophysics.}, \textbf{A13}, 594 (2016).}
	
	\bibitem{Aghanim2020} N. Aghanim et al., \href{https://doi.org/10.1051/0004-6361/201833880}{\textit{Astronomy \& Astrophysics}, \textbf{A1}, 641 (2020).}

        \bibitem{Aghanim2020a} N. Aghanim et al., \href{https://doi.org/10.1051/0004-6361/201833885}{\textit{Astronomy \& Astrophysics}, \textbf{A12}, 641 (2020).}

        \bibitem{Ruiz-Lapuente2010} P. Ruiz-Lapuente, Dark Energy: Observational and Theoretical Approaches (Cambridge University Press, 2010).

        \bibitem{Amendola2010} L. Amendola and S. Tsujikawa, Dark Energy: Theory and Observations (Cambridge University Press, 2010).
          
	\bibitem{Aldrovandi2013} R. Aldrovandi and J.G. Pereira, Teleparallel Gravity (Springer, Dordrecht, 2013), Vol. 173.

 \bibitem{Maluf2001} J.W. Maluf and J.F. da Rocha-Neto, \href{https://doi.org/10.1103/PhysRevD.64.084014}{\textit{Physical Review D}, \textbf{64}, 084014 (2001).}

  \bibitem{Ferraro2008} R. Ferraro and F. Fiorini, \href{https://doi.org/10.1103/PhysRevD.78.124019}{\textit{Physical Review D}, \textbf{78}, 124019 (2008).}

\bibitem{Bengochea2009} G.R. Bengochea and R. Ferraro, \href{https://doi.org/10.1103/PhysRevD.79.124019}{\textit{Physical Review D}, \textbf{79}, 124019 (2009).}

\bibitem{Linder2010} E. V. Linder, \href{https://doi.org/10.1103/PhysRevD.81.127301}{\textit{Physical Review D}, \textbf{81}, 127301 (2010).}
	
	\bibitem{Nester1999} J.M. Nester et al., \href{https://doi.org/10.48550/arXiv.gr-qc/9809049}{\textit{Chinese Journal of Physics}, \textbf{37}, 113 (1999).}
	
	\bibitem{Altschul2015} B. Altschul et al., \href{https://doi.org/10.1016/j.asr.2014.07.014}{\textit{Advances in Space Research}, \textbf{55}, 501 (2015).}
	
	\bibitem{Jimenez2018} J.B. Jimenez et al., \href{https://doi.org/10.1103/PhysRevD.98.044048}{\textit{Physical Review D}, \textbf{98}, 044048 (2018).}

 \bibitem{Subramaniam2023} G. Subramaniam et al., \href{https://doi.org/10.1002/prop.202300038}{\textit{Fortschritte der Physik}, \textbf{71}, 8 (2023).}


\bibitem{Shabani2023} H. Shabani et al., \href{https://doi.org/10.1140/epjc/s10052-023-11722-5}{\textit{European Physics Journal C}, \textbf{83}, 535 (2023).}

\bibitem{Paliathanasis2023} A. Paliathanasis, \href{https://doi.org/10.1016/j.dark.2023.101255}{\textit{Physics of the Dark Universe}, \textbf{41}, 101255 (2023).}
 	
	\bibitem{Dimakis2022} N. Dimakis et al., \href{https://doi.org/10.1103/PhysRevD.106.043509}{\textit{Physical Review D}, \textbf{106}, 043509 (2022).}

\bibitem{Heisenberg2023} L. Heisenberg et al., \href{https://doi.org/10.1140/epjc/s10052-023-11462-6}{\textit{European Physics Journal C}, \textbf{83}, 315 (2023).}

\bibitem{Shabani2023a} H. Shabani et al., \href{https://doi.org/10.1140/epjc/s10052-024-12582-3}{\textit{European Physics Journal C}, \textbf{84}, 285 (2024).}

\bibitem{Subramaniam2023a}G. Subramaniam et al., \href{https://doi.org/10.1016/j.dark.2023.101243}{\textit{Physics of the Dark Universe}, \textbf{41}, 101243 (2023).}
	
	\bibitem{Lu2019} J. Lu et al., \href{https://doi.org/10.1140/epjc/s10052-019-7038-3}{\textit{European Physics Journal C}, \textbf{79}, 530 (2019).}
	
	\bibitem{Lazkoz2019} R. Lazkoz et al., \href{https://doi.org/10.1103/PhysRevD.100.104027}{\textit{Physical Review D}, \textbf{100}, 104027 (2019).}
	
	\bibitem{Jimenez2020} J.B. Jimenez et al., \href{https://doi.org/10.1103/PhysRevD.101.103507}{\textit{Physical Review D}, \textbf{101}, 103507 (2020).}
	
	\bibitem{Ayuso2021} I. Ayuso et al., \href{https://doi.org/10.1103/PhysRevD.103.063505}{\textit{Physical Review D}, \textbf{103}, 063505 (2021).}
	
	\bibitem{Esposito2022} F. Esposito et al., \href{https://doi.org/10.1103/PhysRevD.105.084061}{\textit{Physical Review D}, \textbf{105}, 084061 (2022).}

	\bibitem{Hu2022} K. Hu et al., \href{https://doi.org/10.1103/PhysRevD.106.044025}{\textit{Physical Review D}, \textbf{106}, 044025 (2022).}
	
	\bibitem{Khyllep2021} W. Khyllep et al., \href{https://doi.org/10.1103/PhysRevD.103.103521} {\textit{Physical Review D}, \textbf{103}, 103521 (2021).}
	
	\bibitem{Khyllep2023} W. Khyllep et al., \href{https://doi.org/10.1103/PhysRevD.107.044022}{\textit{Physical Review D}, \textbf{107}, 044022 (2023).}
	
	\bibitem{Sahlu2022} S. Sahlu et al., \href{https://doi.org/10.48550/arXiv.2206.02517}{\textit{arXiv:2206.02517}, (2022).}
	
	\bibitem{Soudi2019} I. Soudi et al., \href{https://doi.org/10.1103/PhysRevD.100.044008}{\textit{Physical Review D}, \textbf{100}, 044008 (2019).}
	
	\bibitem{Barros2020} B.J. Barros et al., \href{https://doi.org/10.1016/j.dark.2020.100616}{\textit{Physics of the Dark Universe}, \textbf{30}, 100616 (2020).}
	
	\bibitem{Anagnostopoulos2021} F.K. Anagnostopoulos et al., \href{https://doi.org/10.1016/j.physletb.2021.136634}{\textit {Physics Letters B}, \textbf{822}, 136634 (2021).}
	
	\bibitem{Atayde2021} L. Atayde et al., \href{https://doi.org/10.1103/PhysRevD.104.064052}{\textit{Physical Review D}, \textbf{104}, 064052 (2021).}
	
	\bibitem{Frusciante2021} N. Frusciante, \href{https://doi.org/10.1103/PhysRevD.103.044021}{\textit{Physical Review D}, \textbf{103}, 044021 (2021).}
	
	\bibitem{Anagnostopoulos2023} F.K. Anagnostopoulos et al., \href{https://doi.org/10.1140/epjc/s10052-023-11190-x}{\textit{European Physics Journal C}, \textbf{83}, 58 (2023).}
	
	\bibitem{Narawade2022a} S.A. Narawade et al., \href{https://doi.org/10.1016/j.dark.2022.101020}{\textit{Physics of the Dark Universe}, \textbf{36}, 101020 (2022).}

	\bibitem{Koussour2022a} M. Koussour et al., \href{https://doi.org/10.1088/1361-6382/ac8c7d}{\textit{Classical and Quantum Gravity}, \textbf{39}, 195021 (2022).,} \href{https://doi.org/10.1016/j.aop.2022.169092}{\textit{Annals of Physics}, \textbf{445}, 169092 (2022).}
	
	\bibitem{Maurya2022} S.K. Maurya et al., \href{https://doi.org/10.1002/prop.202200061}{\textit{Progress of Physics}, \textbf{70}, 2200061 (2022).}
	
	\bibitem{Harko2018} T. Harko et al., \href{https://doi.org/10.1103/PhysRevD.98.084043}{\textit{Physical Review D}, \textbf{98}, 084043 (2018).}
	
	\bibitem{Bahamonde2022} S. Bahamonde et al., \href{https://doi.org/10.1088/1475-7516/2022/08/082}{\textit{Journal of Cosmology and Astroparticle Physics}, \textbf{08}, 082 (2022).}
	
	\bibitem{Capozziello2022} S. Capozziello et al., \href{https://doi.org/10.1016/j.physletb.2022.137229}{\textit{Physics Letters B}, \textbf{832}, 137229 (2022).}

        \bibitem{Bajardi2020} F. Bajardi et al., \href{https://doi.org/10.1140/epjp/s13360-020-00918-3}{\textit{European Physics Journal Plus}, \textbf{135}, 912 (2020).}

        \bibitem{Bajardi2021} F. Bajardi and S. Capozziello, \href{https://doi.org/10.1142/S0219887821400028}{\textit{International Journal of Geometric Methods in Modern Physics}, \textbf{18}, 2140002 (2021).}

        \bibitem{Bajardi2023} F. Bajardi and S. Capozziello, \href{https://doi.org/10.1140/epjc/s10052-023-11703-8}{\textit{European Physics Journal C}, \textbf{83}, 531 (2023).}

        \bibitem{Vagnozzi2020} S. Vagnozzi, \href{https://doi.org/10.1103/PhysRevD.102.023518}{\textit{Physical Review D}, \textbf{102}, 023518 (2020).}

        \bibitem{Adil2023} S.A. Adil et al., \href{https://doi.org/10.1088/1475-7516/2023/10/072}{\textit{Journal of Cosmology and Astroparticle Physics}, \textbf{10}, 072 (2023).}

        \bibitem{Escamilla2023} L.A. Escamilla et al., \href{https://doi.org/10.1088/1475-7516/2024/05/091}{\textit{Journal of Cosmology and Astroparticle Physics}, \textbf{05}, 091 (2024).}

         \bibitem{Bernal2016} J.L. Bernal et al., \href{https://doi.org/10.1088/1475-7516/2016/10/019}{\textit{Journal of Cosmology and Astroparticle Physics}, \textbf{10}, 019 (2016).}

         \bibitem{Adi2021} T. Adi and E.D. Kovetz, \href{https://doi.org/10.1103/PhysRevD.103.023530}{\textit{Physical Review D}, \textbf{103}, 023530 (2021).}

         \bibitem{Odintsov2021} S.D. Odintsov et al., \href{https://doi.org/10.1016/j.nuclphysb.2021.115377}{\textit{Nuclear Physics B}, \textbf{966}, 115377 (2021).}

         \bibitem{Briffa2022} R. Briffa et al., \href{https://doi.org/10.1140/epjp/s13360-022-02725-4}{\textit{European Physics Journal C}, \textbf{137}, 532 (2022).}
        
        \bibitem{Jarv2018} L. Jarv et al., \href{https://doi.org/10.1103/PhysRevD.97.124025}{ \textit{Physical Review D}, \textbf{97}, 124025 (2018).}
	
	\bibitem{Runkla2018} M. Runkla and O. Vilson, \href{https://doi.org/10.1103/PhysRevD.98.084034}{\textit{Physical Review D}, \textbf{98}, 084034 (2018).}
	
	\bibitem{Xu2019} Y. Xu et al., \href{https://doi.org/10.1140/epjc/s10052-019-7207-4}{\textit{European Physics Journal C}, \textbf{79}, 708 (2019).}
		
	\bibitem{D'Ambrosio2022} F. D'Ambrosio et al., \href{https://doi.org/10.1103/PhysRevD.105.024042}{\textit{Physical Review D}, \textbf{105}, 024042 (2022).}
			
	\bibitem{Albuquerque2022} I.S. Albuquerque and N. Frusciante, \href{https://doi.org/10.1016/j.dark.2022.100980}{\textit{Physics of the Dark Universe}, \textbf{35}, 100980 (2022).}
		
	\bibitem{Narawade2022b} S.A. Narawade and B. Mishra, \href{https://doi.org/10.1002/andp.202200626}{\textit{Annalen der Physik}, \textbf{535}, 2200626 (2023).}

        \bibitem{Koussour:2023rly} M. Koussour and A. De,
    \href{https://doi.org/10.1140/epjc/s10052-023-11547-2}{\textit{European Physics Journal C}, \textbf{83}, 400 (2023).}
	
	\bibitem{Ortin2004} T. Ortin, \href{https://doi.org/10.1017/CBO9780511616563}{\textit{Gravity and Strings} (2004).}

       \bibitem{Golovnev2017} A. Golovnev, \href{https://doi.org/10.1088/1361-6382/aa7830}{\textit{Classical and Quantum Gravity}, \textbf{34}, 145013 (2017).}

       \bibitem{Jimenez2018a} J.B. Jimenez et al., \href{https://doi.org/10.1088/1475-7516/2018/08/039}{\textit{Journal of Cosmology and Astroparticle Physics}, \textbf{08}, 039 (2018).}
	
	\bibitem{Amanullah2010} R. Amanullah et al., \href{https://doi.org/10.1088/0004-637X/716/1/712}{\textit{The Astrophysical Journal}, \textbf{716}, 712 (2010).}

 
       \bibitem{Sahni2003a} V. Sahni and Y. Shtanov, \href{https://doi.org/10.1088/1475-7516/2003/11/014}{\textit{Journal of Cosmology and Astroparticle Physics}, \textbf{11}, 014 (2003).}
	
	\bibitem{Capozziello2008} S. Capozziello et al., \href{https://doi.org/10.1016/j.physletb.2008.04.061}{\textit{Physics Letters B}, \textbf{664}, 12 (2008).}

    \bibitem{Yang2020} Y. Yang and Y. Gong, \href{https://doi.org/10.1088/1475-7516/2020/06/059}{\textit{Journal of Cosmology and Astroparticle Physics}, \textbf{06}, 059 (2020).}

    \bibitem{Capozziello2014} S. Capozziello et al., \href{http://dx.doi.org/10.1103/PhysRevD.90.044016}{\textit{Physical Review D}, \textbf{90}, 044016 (2014).}
	
	\bibitem{Sahni2003} V. Sahni et al., \href{https://doi.org/10.1134/1.1574831}{\textit{Journal of Experimental and Theoretical Physics Letters}, \textbf{77}, 201 (2003).}

         \bibitem{Zhang2005} X. Zhang, \href{https://doi.org/10.1142/S0218271805007243}{\textit{International Journal of Modern Physics D}, \textbf{14}, 1097 (2005).}
 
	\bibitem{Wang2009} F. Y. Wang et al., \href{https://doi.org/10.1051/0004-6361/200911998}{\textit{Astronomy and Astrophysics}, \textbf{507}, 53 (2009).}

    \bibitem{Moresco2022} M. Moresco et al., \href{https://doi.org/10.1007/s41114-022-00040-z}{\textit{Living Reviews in Relativity}, \textbf{25}, 6 (2022).}
 
	\bibitem{Brout2022} D. Brout et al., \href{https://doi.org/10.3847/1538-4357/ac8e04}{\textit{The Astrophysical Journal}, \textbf{938}, 110 (2022).}
	
	\bibitem{Scolnic2018} D.M. Scolnic et al., \href{https://doi.org/10.3847/1538-4357/aab9bb}{\textit{The Astrophysical Journal}, \textbf{859}, 101 (2018).}
	
	\bibitem{Giostri2012} R. Giostri et al., \href{https://doi.org/10.1088/1475-7516/2012/03/027}{\textit{Journal of Cosmology and Astroparticle Physics}, \textbf{03}, 027 (2012).}
	
	\bibitem{Komatsu2009} E. Komatsu et al., \href{https://doi.org/10.1088/0067-0049/180/2/330}{\textit{The Astrophysical Journal Supplement Series}, \textbf{180}, 330 (2009).}
	
	\bibitem{Alam2003} U.Alam et al., \href{https://doi.org/10.1046/j.1365-8711.2003.06871.x}{\textit{Monthly Notices of the Royal Astronomical Society}, \textbf{344}, 1057 (2003).}
	
	\bibitem{Ding2015} X. Ding et al., \href{https://dx.doi.org/10.1088/2041-8205/803/2/L22}{\textit{The Astrophysical Journal Letters}, \textbf{803}, L22 (2015).}
	
	\bibitem{Zheng2016} X. Zheng et al., \href{https://dx.doi.org/10.3847/0004-637X/825/1/17}{\textit{The Astrophysical Journal}, \textbf{825}, 17 (2016).}
	
	\bibitem{Qi2018} J-Z. Qi et al., \href{https://dx.doi.org/10.1088/1674-4527/18/6/66}{\textit{Research in Astronomy and Astrophysics}, \textbf{18}, 066 (2018).}
 	
	\bibitem{Sahni2008} V. Sahni et al., \href{https://link.aps.org/doi/10.1103/PhysRevD.78.103502}{\textit{Physical Review D}, \textbf{78}, 103502 (2008).}


	\bibitem{Koussour2022b} M. Koussour et al., \href{https://doi.org/10.1142/S0218271822501152}{\textit{International Journal of Modern Physics D}, \textbf{31}, 16 (2022).}
	
\bibitem{Novello2008} M. Novello et al., \href{https://doi.org/10.1016/j.physrep.2008.04.006}{\textit{Physics Reports}, \textbf{463}, 127 (2008).}	

\bibitem{Carroll2003} S.M. Carroll et al., \href{https://doi.org/10.1103/PhysRevD.68.023509}{\textit{Physical Review D}, \textbf{68}, 023509 (2003).}

\bibitem{Capozziello2019} S. Capozziello et al., \href{https://doi.org/10.1142/S0218271819300167}{\textit{International Journal of Modern Physics D}, \textbf{28}, 1930016 (2019).}

\end{thebibliography}
\end{document}